\documentclass{aa}
\let\linenumbers\relax

\usepackage{graphicx}
\usepackage{tabularx}
\usepackage{txfonts}
\usepackage{xcolor}
\usepackage{amsmath}
\usepackage{url}
\expandafter\def\expandafter\UrlBreaks\expandafter{\UrlBreaks\do\a\do\b\do\c\do\d\do\e\do\f\do\g\do\h\do\i\do\j\do\k\do\l\do\m\do\n\do\o\do\p\do\q\do\r\do\s\do\t\do\u\do\v\do\w\do\x\do\y\do\z\do\-}
\usepackage[normalem]{ulem}

\newcommand{\spfdu}{dB(W\,m$^{-2}$\,Hz$^{-1}$)}
\newcommand{\pfdu}{dB(W\,m$^{-2}$\,MHz$^{-1}$)}

\defcitealias{fcc2024da1193}{FCC DA~24-1193}
\defcitealias{fcc2026da391}{FCC DA~26-391}
\defcitealias{lynk2022fcc}{FCC DA~22-969}
\defcitealias{fcc2024scs}{FCC~24-28}
\defcitealias{ecc2011rep171}{ECC Report~171}
\defcitealias{ecc2026rep373}{ECC Report~373}
\defcitealias{ecc2025rep363}{ECC Report~363}
\defcitealias{ecc2018rep271}{ECC Report~271}
\defcitealias{ecc2020rep309}{ECC Report~309}
\defcitealias{ecc2026dec2207}{ECC Decision (22)07}
\defcitealias{itu2003ra769}{Recommendation ITU-R RA.769}
\defcitealias{itu2024sm329}{Recommendation ITU-R SM.329}
\defcitealias{itu2024sm1539}{Recommendation ITU-R SM.1539}
\defcitealias{itu2024sm1541}{Recommendation ITU-R SM.1541}
\defcitealias{itu2007m1583}{Recommendation ITU-R M.1583}
\defcitealias{itu2007s1586}{Recommendation ITU-R S.1586}
\defcitealias{itu2015ra1513}{Recommendation ITU-R RA.1513}
\defcitealias{itu2018sm2421}{Report ITU-R SM.2421}
\defcitealias{itu2025ra2551}{Report ITU-R RA.2551}
\defcitealias{itu2003ra1631}{Recommendation ITU-R RA.1631}
\defcitealias{itu2026rxresilience}{Report ITU-R RA.[RX-Resilience]}
\defcitealias{itu2023res253}{Resolution 253 (WRC-23)}
\defcitealias{3gpp36104}{3GPP TS~36.104}
\defcitealias{3gpp38108}{3GPP TS~38.108}
\defcitealias{itu2026wp4c35th}{ITU-R WP~4C Document 4C/744}
\defcitealias{acma2013auction}{ACMA 2013 auction summary}
\defcitealias{acma2025esl4}{ACMA 2025}

\begin{document}

   \title{Harmonic and intermodulation emission from direct-to-device satellites}

   \subtitle{Harmonics of the Starlink 2.6\,GHz downlink and the products around them}

   \author{B. Indermuehle\inst{1}\corrauth{balt.indermuehle@csiro.au}
        \and B. Winkel\inst{2}\email{bwinkel@mpifr-bonn.mpg.de}
        \and F. Di Vruno\inst{3}\email{federico.divruno@skao.int}
        \and G. I. G. J\'ozsa\inst{2,4}\email{gjozsa@mpifr-bonn.mpg.de}
        \and L. Louren\c{c}o\inst{1}\email{liroy.lourenco@csiro.au}
        }

   \institute{CSIRO Space \& Astronomy, PO Box 76, Epping NSW 1710, Australia
   \and Max-Planck-Institut f\"ur Radioastronomie, Auf dem H\"ugel 69, 53121 Bonn, Germany
   \and SKAO, Square Kilometre Array Observatory, Jodrell Bank, Cheshire, SK11 9FT, United Kingdom
   \and Centre for Radio Astronomy Techniques and Technologies (RATT), Department of Physics and Electronics, Rhodes University, Makhanda 6140, South Africa}

   \date{Received ...}

  \abstract
   {Direct-to-device (DTD) satellites carry mobile-network base stations transmitting in terrestrial mobile bands. Their non-linear emission products can fall into radio astronomy bands, including passive bands.}
   {We measure the harmonics of the Starlink DTD downlink at 2620--2630\,MHz and the intermodulation products around them, search for products next to the carriers, and compare them with the ITU-R unwanted-emission limits and radio astronomy protection criteria.}
   {We use autocorrelations of single Australia Telescope Compact Array antennas on tracked passes of DTD and other satellites with the 16\,cm and 4\,cm receivers. Harmonic ratios and receiver tests show that the harmonics are satellite-generated. We simulate the aggregate equivalent power flux density of a constellation at Effelsberg and the ATCA.}
   {The harmonics H2 to H4 at 5240, 7860 and 10480\,MHz fall by about 16\,dB per order, show no trend with elevation, and originate on the satellite. H2 is also detected in 32\% of integrations more than 10$^\circ$ from any DTD satellite. Inside the harmonic zones the spectra follow the $n$-fold convolution of the carriers; outside them, intermodulation products of order $n+2$ and higher lie on average 19\,dB below the zone and switch between $-23$ and $-15$\,dB within seconds. Next to the carriers no satellite emission is detected to 33\,dB below the carriers. H2, $\geq -38.9$\,dBW in 4\,kHz isotropic-equivalent, is at or above the SM.329 space-station limit; H3 and H4 meet it. For 8640 satellites the aggregate exceeds the protection criteria by 65 and 25\,dB at the second and fifth harmonics, and by $>$100\,dB for out-of-band emission at the mask level in 2690--2700\,MHz.}
   {Harmonics are received far outside the serving beams. Their radiated level must be limited directly, not through the power at the antenna port. Compliance with per-satellite ITU-R limits does not protect radio astronomy from the aggregate of a constellation.}


   \maketitle

\section{Introduction}\label{sec:intro}

Direct-to-device (DTD) satellites carry base stations for mobile-phone networks and are mostly operating in fleets of large satellite constellations in low Earth orbit. They currently transmit LTE (4G) and in the near future NR-NTN (5G) downlinks in terrestrial mobile bands directly to off-the-shelf mobile devices. SpaceX operates by far the largest fleet. More than 600 of their Starlink satellites are DTD capable \citep[e.g.][]{indermuehle2026emlimits,mcdowell2026starlink} and are authorised by the USA Federal Communications Commission (FCC) in shells at 340--360\,km and for operation outside the United States of America, subject to national authorisation \citepalias{fcc2024da1193}, using downlinks between 1475 and 2690\,MHz. AST SpaceMobile and Lynk both currently have small test fleets only with 5 and 12 satellites respectively, will transmit below 960\,MHz \citepalias{fcc2026da391,lynk2022fcc} and have not been measured yet. A deployment typically reuses one 5\,MHz LTE carrier of a partner operator having a license for terrestrial frequencies, for example 1990--1995\,MHz in the United States (\citetalias{fcc2024da1193}; \citealt{garciacabeza2025}). The harmonics of the DTD downlinks can fall into radio astronomy and geodetic VLBI bands \citep{indermuehle2026sniffles,indermuehle2026emlimits}. DTD is also on the agenda of the next World Radiocommunication Conference (WRC-27, agenda item 1.13), framed by Resolution~253 (WRC-23),  which notes that spurious-domain emissions may be considered with regard to radio astronomy allocations \citepalias[\textit{recognizing}~f;][]{itu2023res253}. \citet{indermuehle2026emlimits} describe this regulatory framework and the ITU-R terminology in more detail.

Non-linear products of airborne and satellite transmitters and their effect on radio astronomy are well studied. \citetalias{ecc2011rep171} traced Iridium interference in the 1610.6--1613.8\,MHz band, allocated to the radio astronomy service, to the seventh-order product $4f_1-3f_2$ generated on the satellite, with the third and fifth orders absent. \citet{jessner2013} reports an EIRP of about 1\,mW, roughly 40\,dB below the carrier, yet single transits exceeded the radio astronomy protection thresholds provided in \citetalias{itu2003ra769} by more than 20\,dB \citepalias[see][]{ecc2011rep171}. The Iridium case is the only one we know of where \emph{intermodulation} products were a subject in spectrum management processes related to radio astronomy. \emph{Harmonic} emissions, on the other hand, are much more often considered. \citetalias{itu2018sm2421} contains measurements of LTE800 user equipment with a clear detection of second harmonics at about 48.5\,dB below the carrier signal. In Europe, \citetalias{ecc2026dec2207} harmonised the use of such user equipment on board unmanned aircraft and established no-transmit zones for the protection of incumbent services including radio astronomy. The second harmonics of LTE700 and LTE800 were considered in this decision, as radio telescopes are particularly vulnerable to air- and spaceborne interference \citepalias[compare Radio Regulations, RR, No.~5.149; compatibility studies can be found in][]{ecc2020rep309}. Further cases of harmonics observed with radio astronomy receivers are compiled in \citetalias{itu2025ra2551}. Examples for harmonics-related provisions in the RR are mobile-satellite service downlinks operating at 2483.5--2500\,MHz, which can impact radio astronomy at 4990--5000\,MHz (RR No.~5.402) and the operation of HIBS\footnote{High Altitude Platform Stations as IMT Base Stations} in the LTE700 and LTE800 bands (RR No.~5.314A). Consequently, \citetalias{ecc2026rep373} (Table~9) lists the ''primary'' RAS band 1610.6--1613.8\,MHz as potentially affected by second harmonics of LTE800 DTD downlinks. An allocation of a band on a primary basis to a service implies an unrestricted right to protection from harmful interference. 

The SNIFFLES-I survey used the Australia Telescope Compact Array and Mopra telescopes to measure the intended emission, unwanted emission and unintended radiation of low-Earth-orbit satellites. Its first paper covered 1--26\,GHz and reports on the detection of harmonics of the Starlink 2.6\,GHz DTD downlink up to the fourth order \citep{indermuehle2026sniffles}. Here, we report the first calibrated measurements of (i) the harmonics H2 to H4 of the two-carrier Starlink DTD downlink in Australia, recorded simultaneously, (ii) the intermodulation products around these harmonics, and (iii) an upper limit on emission next to the carriers. We compare them with the ITU-R unwanted-emission limits and estimate the aggregate impact of a DTD constellation on radio astronomy with epfd simulations.

In this paper we report on transmitter-generated products only. The satellite emits them, and they reach the telescope within its observing band. A receiver can generate harmonics itself when a strong carrier enters it. Intermodulation products can be generated inside the radio astronomy receiver by strong out-of-band downlinks, modelled in Annex~4 of \citetalias{ecc2018rep271} and discussed in the ITU-R report on the resilience of radio astronomy receivers \citepalias{itu2026rxresilience}. The 4\,cm receiver retains some sensitivity at 2.6\,GHz. We therefore thoroughly tested where the 4\,cm harmonics are generated (Sect.~\ref{sec:rxh2}) with a statistically significant sample of the harmonics power ratios: On tracked DTD passes, the ratios H3/H2 and H4/H2 show no trend over more than three decades of H2, and within passes H3 does not follow H2 as a receiver product would. We conclude that these harmonics are generated on the satellite. A weak H2 of a few Jy, seen far from any DTD satellite, has the same H3/H2 ratio and is the aggregate of the satellites received through the far sidelobes. In addition, the noise diode shows that the carriers do not drive the receiver into the non-linear regime (Sect.~\ref{sec:rxh2}), confirming the finding of satellite-generated harmonics.

Section~\ref{sec:products} summarises how the products arise and which radio astronomy bands they reach, Sect.~\ref{sec:obs} describes the observations and data reduction, and Sect.~\ref{sec:results} presents the measurements. Section~\ref{sec:sm329} compares them with the ITU-R limits, Sect.~\ref{sec:epfd} gives the epfd simulations, and Sect.~\ref{sec:conclusions} concludes.

\section{Harmonic and intermodulation products}\label{sec:products}

A memoryless non-linearity driven by carriers at frequencies $f_i$ with bandwidths $B_i$ produces components at
\begin{equation}
f = \sum_i k_i f_i, \quad N = \sum_i |k_i|, \quad n = \sum_i k_i, \quad B \leq \sum_i |k_i|\,B_i,
\label{eq:im}
\end{equation}
with integer $k_i$. The order $N$ counts the carrier factors in the product, and $n$ is the harmonic zone H$n$ in which it falls, near $n$ times the carrier frequency. For example, $2f_1-f_2$ has $N=3$ and $n=1$ (a near-carrier product), and $3f_1-f_2$ has $N=4$ and $n=2$ (in the H2 zone). A single carrier gives harmonics at $nf$ whose spectrum is the $n$-fold self-convolution of the carrier spectrum (Sect.~\ref{sec:single}). With several carriers, each zone also holds mixed products with multinomial weights (Sect.~\ref{sec:multi}). Sections~\ref{sec:dtd} and \ref{sec:ras} list the DTD bands and the radio astronomy and VGOS bands reached by their harmonic zones.

\subsection{Single carrier}\label{sec:single}

For a memoryless amplifier $y=\sum_n a_n x^n$ driven by a carrier of bandwidth $B$, the $n$-th harmonic lies at $nf$ and has width $nB$. Its spectrum is the $n$-fold self-convolution of the carrier spectrum, which smooths towards a Gaussian as $n$ grows. The $n$-th harmonic rises by $n$\,dB per dB of drive. This drive-level slope describes changes about an operating point, not the level at that point; the level ladder of Sect.~\ref{sec:res:harm} is measured at one operating point.

\subsection{Multiple carriers}\label{sec:multi}

For carriers of equal amplitude the leading term of a product with indices $k_i$ (Eq.~\ref{eq:im}) has power $20\log_{10}(N!/\prod_i |k_i|!)$\,dB relative to the pure $N$-th harmonic of one carrier. Examples: $f_1+f_2$ $+6.0$\,dB, $2f_1+f_2$ $+9.5$\,dB, $f_a+f_b+f_c$ $+15.6$\,dB. These are fixed offsets for equal carriers, not drive-level slopes; a product $af_1+bf_2$ rises $a$\,dB per dB of $f_1$ and $b$\,dB per dB of $f_2$. Near-carrier products ($n=1$) form a comb at multiples of the carrier spacing, and difference products ($n=0$) fall at low frequencies. Cross-band products exist only if several bands share one amplifier; for example, $f_1+f_2$ with carriers in 2345--2360 and 2620--2690\,MHz falls in the radio astronomy band 4990--5000\,MHz. \citetalias{ecc2011rep171} is the measured precedent for a high-order mixed product ($4f_1-3f_2$) reaching a radio astronomy band.

\subsection{Near-carrier products and the place of generation}\label{sec:diag}

Near-carrier products searched for in this paper have $n=1$ and odd $N$. For the two DTD carriers operating at $f_1=2622.5$ and $f_2=2627.5$\,MHz, the third-order product $2f_1-f_2$ (IM3) lies at 2617.5\,MHz and the fifth-order $3f_1-2f_2$ (IM5) at 2612.5\,MHz, with mirror products at 2632.5 and 2637.5\,MHz. At the ATCA the mirror products fall on frequency bands already in use by terrestrial LTE downlinks at 2630.5--2669\,MHz, which are 10 to 100 times stronger, so only the lower-frequency products could be searched for. Section~\ref{sec:res:nearim} gives the result of that search. The dynamic range of the receiving system is not high enough, thus only an an upper limit can be determined. However, the diagnostics below are also applied to the products around the harmonics, which are indeed detected (Sect.~\ref{sec:res:harmim}), and to the harmonics themselves (Sect.~\ref{sec:rxh2}). A product of order $N$, generated in an amplifier or a receiver, rises by $N$\,dB per dB of drive, so its ratio to the carrier rises by $N-1$\,dB per dB. Products generated in the receiving system must be separated from products radiated by the satellite. Products generated in the receiver follow this law in received carrier power. Products radiated by the satellite keep a fixed ratio to the received carrier, whatever the range or telescope gain. Changes of the drive level on the satellite amplifiers would change this ratio, by $N-1$\,dB per dB, and so add scatter to it, but they are not correlated with the range and telescope gain that set the received carrier level; changes of the radiated power at a fixed drive per amplifier, for example in the number of active beams, leave the ratio unchanged. Section~\ref{sec:mitigation} summarises back-off, channel planning, filtering, and array effects, which play a role in the generation of intermodulation products.

The spectral shape of each product gives a second diagnostic, which identifies the carriers as the origin of an emission, but not of the non-linearity, since a satellite and a receiver would produce the same shape. If the carriers are independent, noise-like signals, as OFDM approximately is, a memoryless non-linearity gives an order-$N$ product of the $N$-fold convolution of the carrier spectra, with one spectrum mirrored for each negative $k_i$. The convolution smooths out the fine structure of the carriers and makes each product a smooth hump that falls to zero at the edge of its range. We compare this model with the measured spectra in the harmonic zones (Sect.~\ref{sec:res:harmim}). If the measured profile matches this convolution, the emission in the zone is a product of the carriers, made by a memoryless non-linearity acting on both of them; if it does not, the carriers carry correlated content or the emission has another origin.

\subsection{DTD downlink bands}\label{sec:dtd}

Measured and authorised DTD downlinks include 2620--2630\,MHz in Australia (this work; two LTE 5\,MHz carriers), 1990--1995\,MHz in the United States (\citetalias{fcc2024da1193}; \citealt{garciacabeza2025}), and 1475--1518, 1805--1880, 1930--2000, 2110--2180, 2180--2200, 2345--2360 and 2620--2690\,MHz, the bands in which the FCC authorises SpaceX to operate outside the United States, subject to authorisation by each country \citepalias[para.~39]{fcc2024da1193}. AST SpaceMobile and Lynk use 617--960\,MHz \citepalias{fcc2026da391,lynk2022fcc}. The UK framework covers the FDD and SDL bands 738--758, 758--788, 791--821, 925.1--959.9, 1452--1492, 1805.1--1880, 2110.3--2169.7 and 2620--2690\,MHz \citep[Table~3.1]{ofcom2025d2dsept}, with actual downlink licenses at 925.1--930.1\,MHz and 1805.1--1810.9\,MHz \citep{uksi2026d2d, uksi2026d2damend}. In Europe, DTD is being addressed in different bands: the EU 2\,GHz MSS downlink at 2170--2200\,MHz \citep{ec2026mss} and national authorisation in mobile bands \citepalias{ecc2026rep373}. The bands therefore differ between regions, and the harmonic zones of Sect.~\ref{sec:ras} must be checked for each. Every Starlink DTD deployment we know of uses 5\,MHz LTE carriers; Australia is the only measured two-carrier case.

\subsection{Radio astronomy bands and harmonic zones}\label{sec:ras}

The $n$-th harmonic zone of a band $[f_a,f_b]$ is $[nf_a,nf_b]$. Here are some examples where harmonic zones overlapped with primary RAS bands or VGOS bands: $2\times(791\text{--}821)$\,MHz covers the 1610.6--1613.8\,MHz OH band \citepalias[Table~9]{ecc2026rep373}. $2\times(2620\text{--}2690)$\,MHz falls in VGOS band B (4800--5850\,MHz) and $3\times(2620\text{--}2690)$\,MHz just below the legacy X band \citep{indermuehle2026emlimits}. The radio astronomy band 10.6--10.7\,GHz is reached by $5\times(2120\text{--}2140)$\,MHz, inside the widely used mobile downlink band 2110--2170\,MHz, and by $4\times(2650\text{--}2675)$\,MHz. $5\times(2620\text{--}2690)$\,MHz falls in VGOS band D (12.75--14.0\,GHz). Downlinks in 852--960\,MHz place none of their harmonics 2 to 5 in a geodetic band \citep{indermuehle2026emlimits}. Three cases are studied in Sect.~\ref{sec:epfd}: out-of-band emission of downlinks at the top of 2620--2690\,MHz, next to the passive band 2690--2700\,MHz (RR No.~5.340); the second harmonic of 2483.5--2500\,MHz, allocated to the mobile-satellite service, in 4990--5000\,MHz; and the fifth harmonic of 2120--2140\,MHz in 10.6--10.7\,GHz.

\section{Observations}\label{sec:obs}

\subsection{The BIGCAT backend}

BIGCAT retains the Australia Telescope Compact Array (ATCA)'s Compact Array Broadband Backend (CABB) first IF stage, which brings every receiver to a common 4 to 12 GHz intermediate frequency (IF). The 4\,cm receiver's RF output is directly sampled without downconversion. The mm receivers (15\,mm, 7\,mm and 3\,mm) are mixed with a reference local oscillator (LO) to downconvert their RF into 4 to 12\,GHz, which is why no receiver can observe with an instantaneous bandwidth of more than 8 GHz. The 16\,cm receiver's signal is also going through a mixer: its 1.1 to 3.1\,GHz RF is upconverted to 9.64 to 11.64\,GHz before sampling, and its maximum instantaneous bandwidth is 2\,GHz.

The 4 to 12\,GHz IF is then split, still in the analogue chain, into four 2048\,MHz bands, each with its own LO tunable over 800\,MHz in 160\,MHz steps, so the bands can be butted together for 7.7\,GHz of contiguous coverage or spread apart. Each 2\,GHz wide band and polarisation is digitised separately by one ADC on a Jimble RFSoC board at 4096\,MSps, with two Jimbles per antenna since only four of the eight ADCs on each board are used.

A coarse filterbank runs in the Jimble FPGA. It is an oversampled polyphase filterbank with a 32/27 oversampling ratio that cuts each 2\,GHz band into fifteen subbands spaced 128\,MHz apart but about 152\,MHz wide, so adjacent subbands overlap and there is no roll-off gap. These subbands are requantised to 8 or 16\,bits, framed as CODIF and streamed over 100\,GbE to a GPU cluster.

A fine filterbank is implemented on the GPU. Each GPU stream takes one subband from all six antennas, applies delay and fringe rotation (for interferometry, not relevant to our autocorrelations), and runs a windowed FFT of a few thousand points to produce fine channels, which are cross-correlated and then averaged down to the continuum resolution or kept finer in spectral windows. The 1.157\,kHz channels used here come from BIGCAT's so-called zoom bands, which yield higher spectral resolution over a selected part of a subband.

All data processed in this paper are the autocorrelations of single ATCA antennas, recorded with the BIGCAT back-end in 2.07\,s integrations (0.26\,s in December 2025), with the subbands of its filterbank requantised to 8 bits and channelised to 1.157\,kHz widths, and averaged to 100\,kHz unless stated otherwise. Spectra are the mean of the two linear polarisations unless stated otherwise. The DTD carriers (including surrounding frequencies) and their harmonic features were observed with two different receivers: The 16\,cm receiver covering the frequency range 1.1--3.1\,GHz and the 4\,cm receiver covering the frequency range 3.9--12.2\,GHz

\subsection{Emission next to the carriers} 

The data come from the SNIFFLES campaign \citep{indermuehle2026sniffles}, which tracks satellites of many constellations to characterise their emissions; most passes therefore target non-DTD satellites. We extracted 2556--2812\,MHz spectra from 156 passes, each one tracking a target satellite from the start to the end of its track, observed with the 16\,cm receiver in May and June 2026, in 100\,kHz channels (native 1.157\,kHz): 72 Starlink (11 of them DTD), 62 Hulianwang Digui, 10 OneWeb, 7 Jilin and 4 SpaceMobile satellites and Optus~D3. In addition, 20 calibrator scans were carried out yielding the flux-density scale. Tracked DTD passes saturate the back-end, so the carrier data come from the 165 scans with non-DTD targets: the remaining 145 satellite passes and the 20 calibrator scans, during which DTD satellites also cross the sidelobes. The calibrators add at most a few per cent of the system noise, a steady continuum that the baseline subtraction removes, which means that they do not differ from the non-DTD scans for the purposes of our analyses. In all these scans, DTD satellites sometimes come close to the main beam or cross near sidelobes, but the carrier usually stays within the linear range of the back-end. During beam transits, 1--2 integrations saturate it; we discarded them. Antenna CA01 observed 34 passes on 22--25 May and CA02 122 passes and the 20 calibrator scans on 2--26 June. We use only the June data from CA02, 140 scans with non-DTD targets (120 passes and 20 calibrator scans): the May data from CA01 carry about 2.7 times stronger terrestrial LTE and an 8 times noisier reference band.

The Australian DTD downlink consists of two 5\,MHz LTE carriers at 2622.5 and 2627.5\,MHz. To remove the system bandpass (the frequency-dependent gain of the receiving system), for every 2.07\,s integration of a scan we formed $r=P/P_0-1$ per channel, with $P_0$, for each channel, the 10th percentile of its values over all integrations of the scan, which estimates the bandpass from the quietest tenth of the scan. We then subtracted the mean $r$ of the integrations with the weakest DTD carrier (below 0.5 of the system noise) to account for mild temporal baseline variations\footnote{Especially at low elevations, the system temperature baseline level changes owing to different amounts of atmospheric noise and ground spillover radiation entering the receiver.}, which also removes steady terrestrial signals (including terrestrial LTE carriers above the DTD band). The carrier level is the mean $r$ over 2621--2629\,MHz. We divided each integration by its own carrier level, so the product-to-carrier ratio is formed per integration, and averaged these normalised spectra in 5\,dB bins of received carrier level, 5--10, 10--15 and 15--20\,dB above the system noise. We call such an average of integrations, each normalised to its own signal level, a stack; 0\,dBc in a stack is the mean in-block power density of the carriers. Most integrations are sidelobe crossings with unknown telescope gain towards the satellite, so these levels are neither pfd nor EIRP of the satellite. The analysis uses only the ratio of products to carrier, which is independent of gain and range and needs no flux density calibration. The flux density of the fundamental on a tracked DTD pass (Sect.~\ref{sec:res:harm}) is taken from \citet{indermuehle2026emlimits}.

\begin{figure}
\centering
\includegraphics[width=\columnwidth]{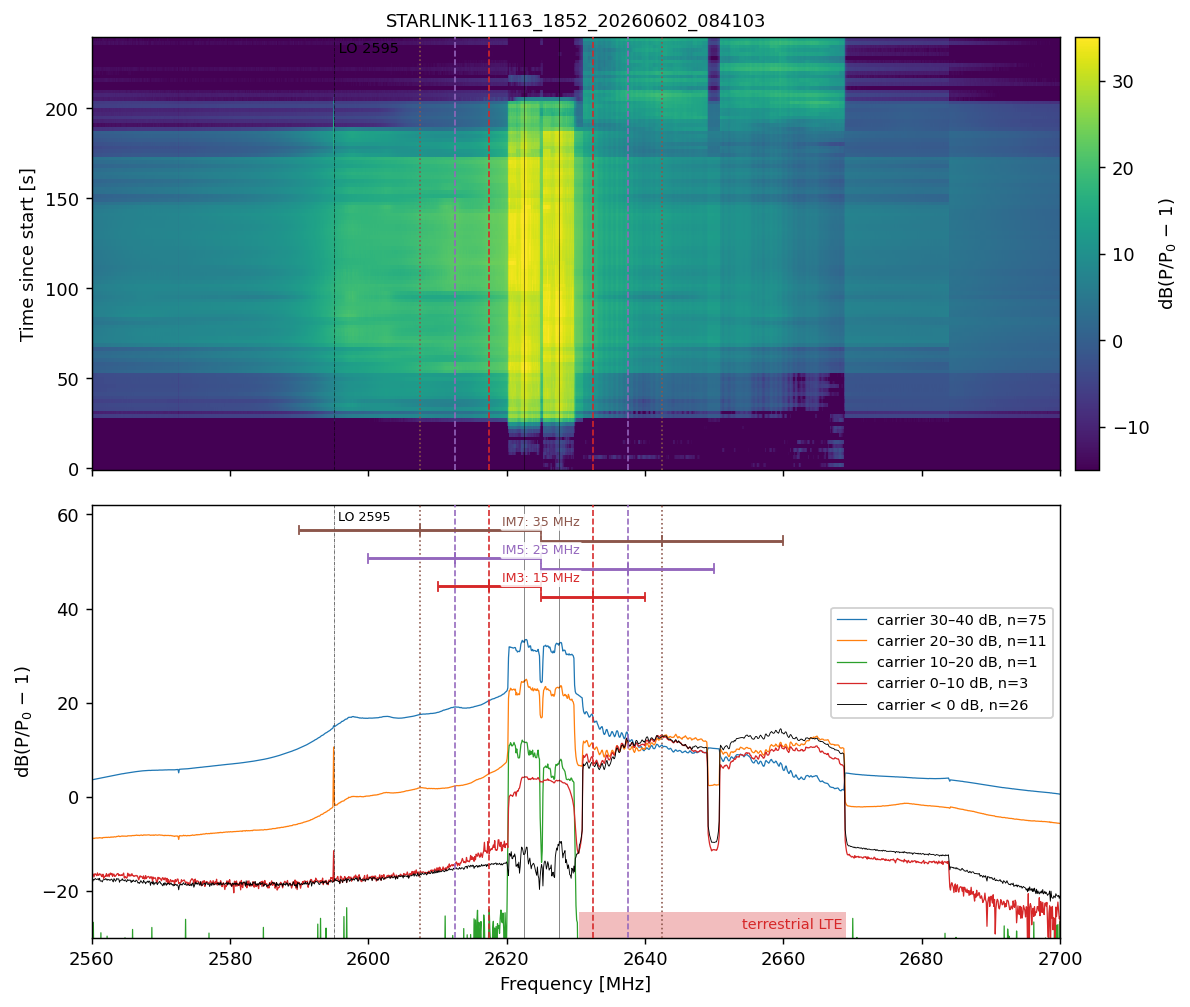}
\caption{Tracked DTD pass STARLINK-11163 (2026-06-02, CA02, 16\,cm). Top: dynamic spectrum, $r$ in dB. Bottom: mean spectra in bins of carrier level (number of integrations in the legend). At carrier levels above 15\,dB the whole band lifts, also above the terrestrial LTE at 2669--2700\,MHz where nothing emits. The step at 2684\,MHz, a digitiser subband boundary, appears in the residual baseline at low carrier levels and is covered by the lift at high levels.}
\label{fig:compression}
\end{figure}

We removed two effects of the receiving system, both visible on the tracked DTD pass STARLINK-11163 (Fig.~\ref{fig:compression}). First, when a DTD satellite passes close to the beam, the strong carrier saturates the requantiser of the back-end subband containing it. This raises the noise floor of that subband and, at the highest levels, lifts the spectrum also at 2720--2790\,MHz in the next subband. We flagged integrations whose median $r$ over 2720--2800\,MHz, i.e., in the adjacent subband, exceeds 0.01 of the system noise, which removed 46 of 2534 integrations. A lift too weak to be flagged is broadband and would raise the empty band 2602--2609,MHz as well; there the stacks of the 5--10 and 10--15\,dB bins stay below $-53$\,dBc, 20\,dB under the level next to the carriers, so a residual lift does not affect the result.  In the 5--10 and 10--15\,dB carrier bins the stacks stay below $-53$\,dBc, 20\,dB under the level next to the carriers. In the 15--20\,dB bin the whole band outside the carriers lies at $-38$ to $-40$\,dBc, the onset of the lift that dominates at higher levels: above a carrier level of about 20\,dB the back-end compresses in the subband that holds the carriers (Sect.~\ref{sec:rxh2}), and the power next to the carriers rises relative to the carrier, to $-23.3$ and $-16.5$\,dBc in the 20--25 and 25--30\,dB bins. Therefore to create a clean data set, only integrations with carrier levels below 15\,dB are kept. 116 scans contribute, with a total of 1925 integrations; six passes whose carrier-free integrations have a disturbed background owing to other types of interference are excluded.

\subsection{Harmonics} 

The 4\,cm receiver recorded 4.7--12.3\,GHz spectra on tracked passes of STARLINK-11xxx DTD satellites with antenna CA02 between 31 May and 3 July 2026. A further 47 passes recorded with CA01 on 24 May 2026 are discarded, since the calibration noise source of that antenna was off and the state of its receiver chain could not be verified. The fundamental lies below this band; H2, H3 and H4 are detected at about 5240, 7860 and 10480\,MHz. Flux densities are $S=\mathrm{SEFD}(f)\,(P/P_0-1)$, with SEFD from a Y-factor table (284, 168 and 114\,Jy at the three frequencies), a 5-channel median filter in frequency for robustness, and where measured at the peak over the integrations of a pass. For H2 this is a lower bound of the real flux density as the back-end compresses for very intense events. We range-corrected the flux densities with $(d/890\,\mathrm{km})^2$, 890\,km being the median range of the tracked passes. For comparison with ITU-R studies, which mostly quote satellite pfd's at nadir, Table~\ref{tab:harmonics} and Fig.~\ref{fig:ladder} give the values at the 350\,km nadir range of the DTD shells.

The ATCA injects a switched calibration noise signal throughout each scan, and the back-end records it separately in every integration, so its amplitude monitors the system gain continuously. This also allows us to monitor whether the analogue and/or digital stages in the system are subject to saturation effects during the measurements. If the noise diode's signal (in arbitrary units), which can be derived from subtracting the data with and without the diode being active, drops, this is an indicator for compression. If all subbands showed this, the likelihood is high that the analogue part of the receiver (the amplifiers or the mixers) are saturated. Likewise, if all subbands associated with one of the four 2-GHz IF blocks would be affected, the ADC could be subject to clipping. Lastly, if just one or few subbands reveal a drop in the noise diode counts, it must be related to the subband processing units and saturation of the analogue part of the receiver can be ruled out.

In the H2 band the normalised noise diode level stays within 0.3\,dB of its nominal value up to about 100\,kJy of recorded satellite signal, then drops to a median of 0.68 ($-1.7$\,dB) at 100--150\,kJy and 0.40 ($-3.9$\,dB) at 150--200\,kJy, with the diode level fluctuating between 0.03 and 0.8 from one integration to the next, as the requantisers of the two back-end subbands that hold the H2 zone compress (knee near 128\,kJy). Where the gain falls below 0.5 the recorded level is a ceiling, and we do not correct it. The adjoining subbands and the H3 and H4 bands do not compress (Sect.~\ref{sec:rxh2}). H2 peaks are therefore lower bounds. With the detection criterion of Sect.~\ref{sec:rxh2}, at least three on-axis integrations above $5\sigma$ with the shape of the zone, H2, H3 and H4 are detected in 45, 29 and 20 of the 55 tracked passes; in 16 of the H2 passes the recorded level exceeds 28\,kJy, where the back-end compresses. Every 4\,cm scan records H2, H3 and H4 at the same time. We call the levels of the successive orders in one pass the harmonic ladder; it isolates the drop from one order to the next, because all orders share the same satellite, range, pointing and drive level. The scan with the strongest signal is a tracked measurement of STARLINK-11117 (2026-05-31 20:52\,UT).

\subsection{Products around the harmonics}

We extracted each harmonic zone $\pm80$\,MHz from the 214 complete 4\,cm observations, in 100\,kHz channels. H2 above the system noise (about 280\,Jy) appears in only 23 integrations of non-DTD passes, too few to stack, so we used tracked DTD passes: 50 with CA02 at tuning 8448\,MHz (31 May, 1 June and 3 July 2026), 10 with CA04 (21 July 2026, H2 only), and 17 with CA05 on 4 December 2025 (0.26\,s integrations), each of which recorded one 1\,GHz\footnote{The December data was recorded with an older firmware where the backend offered only 1 GHz of bandwidth.} block and hence one harmonic (12 at H2, 5 at H3). Only the antenna that tracked the satellite is used; in December 2025, for technical reasons the back-end required a second antenna, CA01, which stayed stowed. The December stacks rest on three scans at H2 and three at H3, a single scan providing 90\% of the integrations at H2 and 86\% at H3. We divided each integration by its own harmonic level, the mean over the central 70\% of the zone (5243--5257, 7864.5--7885.5 and 10486--10514\,MHz for H2, H3 and H4), so 0\,dB is the mean power density in the zone centre. Then we stacked integrations with levels of 10--20\,dB (H2), 5--20\,dB (H3) and 0--10\,dB (H4) above the system noise (Fig.~\ref{fig:harmim}), with these levels chosen per harmonic as the range between the detection threshold of the wings and the onset of compression. H2 integrations above 20\,dB (about 28\,kJy) are excluded because the back-end compresses there. The system noise changes by up to 20\% along a track, which after division by the harmonic level is comparable to the emission outside the zones; we therefore subtracted from each integration a pedestal, the median of eight sub-bands 20--45\,MHz outside the zone edges. Consecutive integrations are correlated, since the transmitter stays in one state for seconds (Sect.~\ref{sec:res:harmim}), so the scatter between single integrations underestimates the error. We therefore estimated the error of each stacked level in two ways: By a bootstrap over passes and from the scatter of the means of contiguous blocks of integrations, and quote the larger of the two.

\section{Results}\label{sec:results}

\subsection{Emission next to the carriers}\label{sec:res:nearim}

In the clean CA02 stack of scans with non-DTD targets, in which the DTD carriers enter through the sidelobes (Sect.~\ref{sec:obs}), the power density in 2610.75--2619.75\,MHz lies 33.1\,dB below that of the carriers ($-32.9$ and $-33.6$\,dBc in the 5--10 and 10--15\,dB carrier bins; 605 integrations at full resolution). Most of this power is instrumental: a skirt of the receiving system that differs from antenna to antenna, rises towards the block edge, and whose origin we cannot determine.

We therefore detect no emission from the satellite next to the carriers. Its power density in 2610.75--2619.75\,MHz is below $-33.1$\,dBc without any modelling. In the third and fourth adjacent channels below the block, 2600--2610\,MHz, the stack lies at $-47$\,dBc, providing an upper bound to the out-of-band emission of the satellite. These limits are set by the dynamic range of the receiving system, so a non-detection is owed to the fact that the DTD signal only enters through the telescope side lobes. Terrestrial LTE base stations must keep 45\,dB of adjacent-channel leakage ratio \citepalias{3gpp36104}, satellite access nodes in low Earth orbit 24\,dB \citepalias[clause~6.6.3.2]{3gpp38108}; the limit lies between the two. Terrestrial LTE at 2630.5--2669\,MHz masks the band above the block.

\subsection{Products around the harmonics}\label{sec:res:harmim}

\begin{figure}
\centering
\includegraphics[width=\columnwidth]{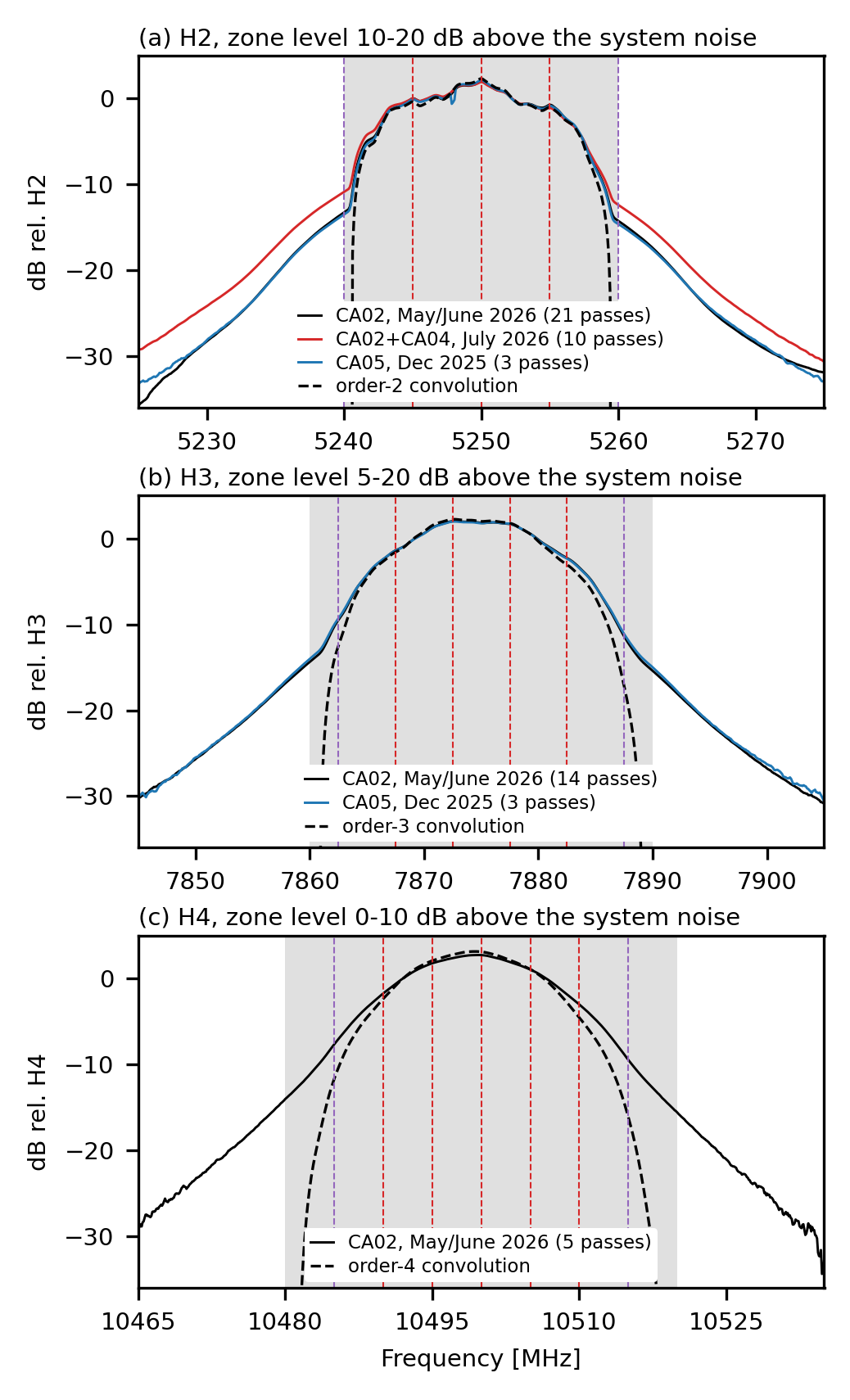}
\caption{Products around H2 (a), H3 (b) and H4 (c) from tracked DTD passes (4\,cm, 100\,kHz), stacked per antenna and epoch, with the pedestal subtracted, for zone levels of 10--20\,dB (H2), 5--20\,dB (H3) and 0--10\,dB (H4) above the system noise. The legend gives the passes contributing to each stack, fewer than the passes with a detection, since only integrations in these level ranges enter. Grey band: the harmonic zone, $n$ times the 10\,MHz carrier block; emission outside it comes from products of order $n+2$ and higher, for example $3f_1-f_2$ at 5240\,MHz. Black dashed: the expected harmonic spectrum, the $n$-fold self-convolution of the measured carrier spectrum, scaled to the zone. Red dashed: order-$n$ product centres; purple dashed: order-$(n+2)$ centres. Beyond about 12\,MHz from the zone edges the curves approach zero, since the level at 20--45\,MHz outside the zone is subtracted from each integration (Sect.~\ref{sec:obs}).}
\label{fig:harmim}
\end{figure}

Figure~\ref{fig:harmim} shows the stacks around H2, H3 and H4. Inside each zone, the profile follows the $n$-fold self-convolution of the measured carrier spectrum down to about $-12$\,dB, with the scale as the only free parameter. At full resolution, the H2 zone shows soft peaks about 2\,MHz wide at the order-2 centres 5245, 5250 and 5255\,MHz, the convolution of the carriers' central plateaus, and no DC notch. The zone ends in a step at 5240.5 and 5259.5\,MHz, twice the carriers' occupied-band edges.

Beyond the zone edges the emission continues. Measured above the pedestal and averaged over 0.5--9.5\,MHz outside both edges, it lies $19.3\pm0.4$\,dB below the mean zone density at H2 (CA02, 31 May and 1 June 2026; 21 passes, 956\,s on the harmonic), $19.8\pm0.3$\,dB at H3 and about 19\,dB at H4, where one pass holds most of the integrations. At H2 it falls from $-15.8$\,dB within 3\,MHz of the edge to $-20.0$\,dB at 3--6\,MHz and $-25.3$\,dB at 6--9.5\,MHz, and the lower wing is about 1\,dB stronger than the upper. The December 2025 data from CA05 give $-19.4\pm0.6$\,dB at H2 and $-19.6\pm0.7$\,dB at H3, and the H3 profiles of the two antennas coincide down to $-30$\,dB, although antenna, epoch, integration time and the position of the zone within the back-end subbands all differ. This sample is small, one dominant pass per harmonic, so it checks the 2026 result rather than repeating it. A third receiver chain adds to it: the parked antenna CA01 saw one December pass by chance at 5--20\,dB above its system noise (11 integrations) and shows the same wings, $-16.4$, $-20.8$ and $-26.8$\,dB at 0.5--3, 3--6 and 6--9.5\,MHz below the zone, 1 to 2\,dB below CA02 in 2026. The wings therefore arrive with the signal and are not an effect of the receiving system.

Products of order $n+2$, such as $3f_1-f_2$ at 5240\,MHz, are the lowest order that falls outside the zone. The zone profile contains the mixed products of both carriers, which shows that the two carriers pass through one non-linearity, and such a non-linearity produces these orders as well. The order-$(n+2)$ convolution alone falls off faster than the measured wings beyond about 5\,MHz from the edge, so higher orders contribute.

\begin{figure*}
\centering
\includegraphics[width=0.95\textwidth]{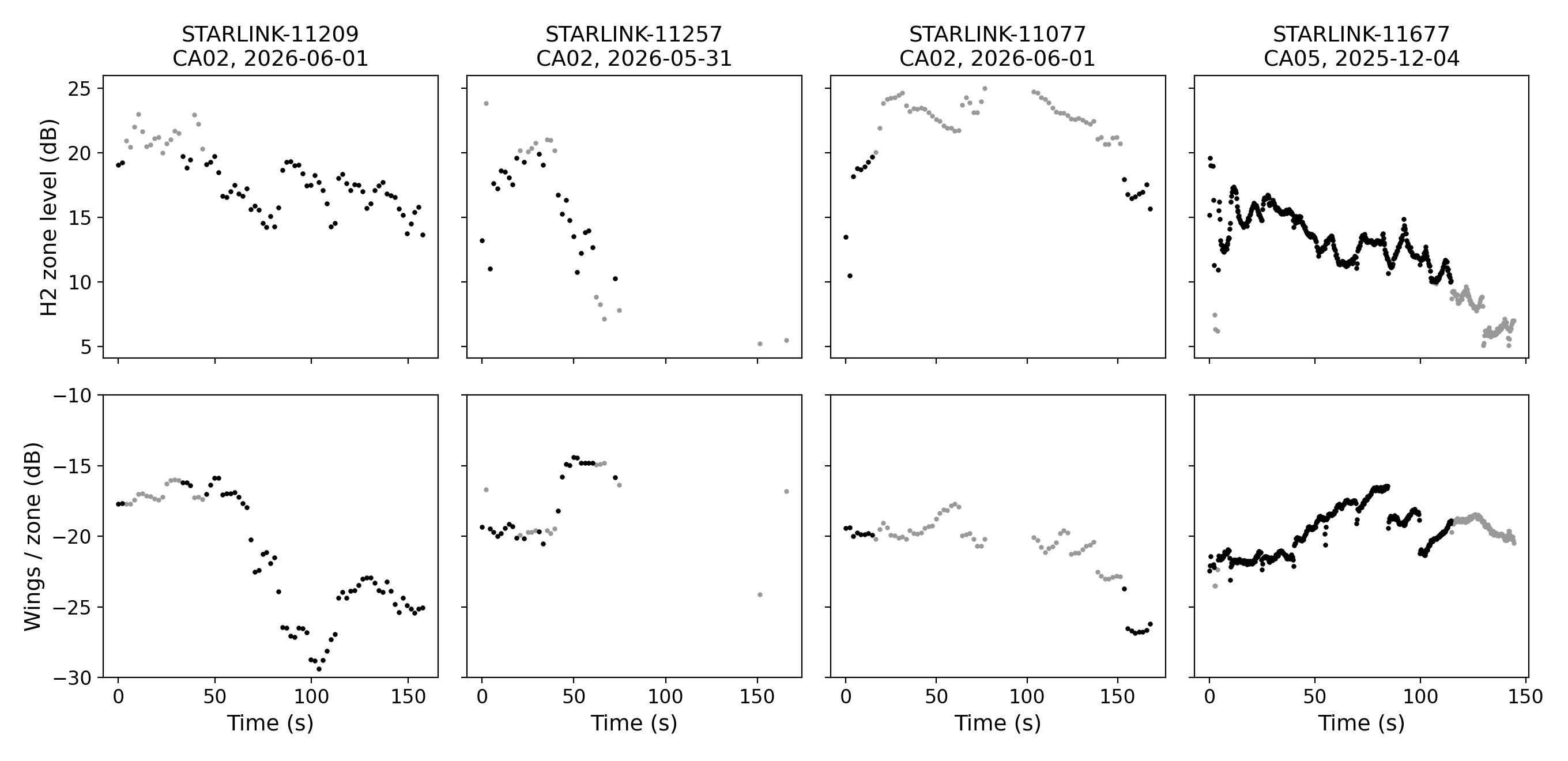}
\caption{H2 zone level (top) and ratio of the wings to the zone (bottom; 0.5--9.5\,MHz outside both zone edges) per integration, for three passes with CA02 in 2026 and one with CA05 in December 2025. Black: zone level 10--20\,dB above the system noise; grey: 5--10 and 20--25\,dB. The level is not corrected for range, which changes by 2 to 3\,dB within these passes; a level of $L$\,dB corresponds to an EIRP of $L + 20\log_{10}(d/890\,\mathrm{km}) - 69.5$\,dBW in 4\,kHz under the assumptions of Sect.~\ref{sec:res:harm}.}
\label{fig:states}
\end{figure*}

The ratio of the wings to the zone is not constant (Fig.~\ref{fig:states}). Within single passes it stays on plateaus and changes by 5 to 9\,dB within one or two integrations; over all integrations at H2 it ranges from $-23$ to $-15$\,dB (5th to 95th percentile). Both wings and both polarisations change together, while the shape of the zone profile, normalised to its own level, changes by less than 1\,dB. Inside passes the ratio follows neither the zone level ($0.03\pm0.10$\,dB per dB) nor H3/H2 ($0.06\pm0.09$\,dB per dB, where a single non-linearity driven harder would give 2), and H3/H2 itself varies by 7 to 16\,dB within a pass. The products around the harmonics thus change with a state of the transmitter that we cannot identify, and not with its drive level alone. The passes of July 2026 (CA02 and CA04) were on average 0.5 to 3.0\,dB higher, depending on the selection, and the difference disappears when passes with a single integration are included; with six to ten passes we cannot distinguish this from the states sampled.

\subsection{Harmonics}\label{sec:res:harm}
\begin{figure}
\centering
\includegraphics[width=\columnwidth]{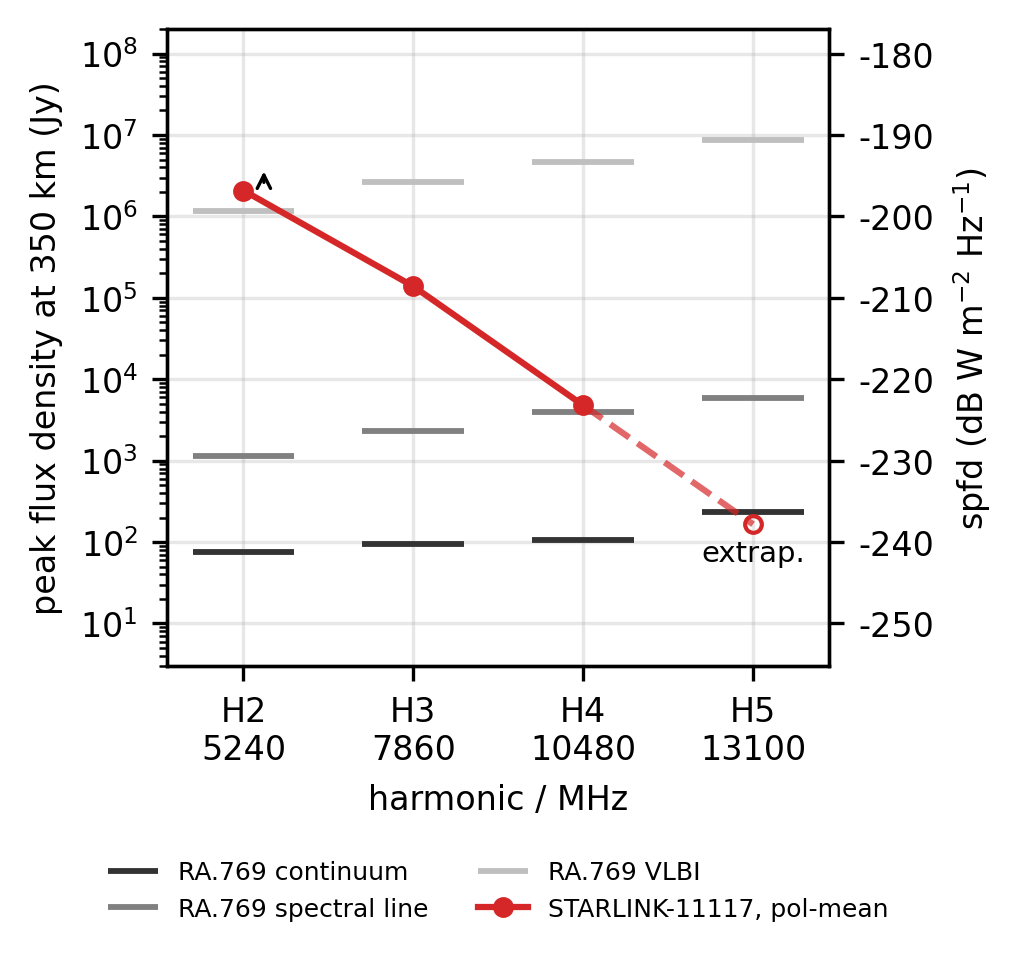}
\caption{Harmonic ladder at 350\,km nadir range of the DTD shells, scaled from the measured 890\,km reference by $(890/350)^2$ ($+8.1$\,dB). Red: STARLINK-11117, mean of the two polarisations. The arrow marks H2 as a compressed lower bound. Open symbols at H5 are extrapolated with the H3$\to$H4 step, not measured. Horizontal bars:  \citetalias{itu2003ra769} thresholds for a single satellite in a 0\,dBi sidelobe, interpolated in frequency between the tabulated bands.}
\label{fig:ladder}
\end{figure}

Table~\ref{tab:harmonics} and Fig.~\ref{fig:ladder} give the harmonics of STARLINK-11117. On this pass the flux density drops by at least 11.8\,dB from H2 to H3 and by 14.6\,dB from H3 to H4; the first value is a lower limit because H2 is compressed. Applying the 14.6\,dB drop to H2 predicts an H2 level 2.8\,dB above the measured value. The noise-diode gain in the ten strongest integrations of this pass is 0.03 to 0.79 (median 0.37, $-4.3$\,dB), so the recorded peak is compressed by 2 to 4\,dB or more, and the H2 to H3 step lies between 11.8\,dB and about 16\,dB. Each harmonic of this pass is therefore about 14--16\,dB fainter than the one before. Over all on-axis integrations of the tracked passes (Sect.~\ref{sec:rxh2}), H3/H2 is $-16.8$\,dB and H4/H2 $-32.9$\,dB, about 16\,dB per order, with 2\,dB scatter between H2 levels; STARLINK-11117 lies at the shallow end. STARLINK-11117 is the strongest of the 55 tracked passes at H3 and H4; at H2, where the strongest passes are compressed, four passes reach similar recorded levels. Extrapolating the H3 to H4 step gives 165\,Jy at 13.1\,GHz for H5 (about $\pm5$\,dB); H5 was not observed.

\begin{table}
\caption{Harmonics of the 2620--2630\,MHz downlink, STARLINK-11117 (2026-05-31 20:52\,UT), H2--H4 recorded simultaneously, mean of the two polarisations, scaled to the 350\,km nadir range of the DTD shells.}
\label{tab:harmonics}
\centering
\small
\setlength{\tabcolsep}{3pt}
\begin{tabular}{rrrrrr}
\hline\hline
$n$ & $f$ & $S$ & spfd & EIRP$_{4\,\mathrm{kHz}}$ & vs.\ SM.329 \\
 & (MHz) & (Jy) & \spfdu & (dBW) & (dB) \\
\hline
2 & 5240 & $\geq2{,}080{,}400$ & $\geq-196.8$ & $\geq-38.9$ & $\geq+1.1$ \\
3 & 7860 & $138{,}500$ & $-208.6$ & $-50.7$ & $-10.7$ \\
4 & 10480 & $4{,}780$ & $-223.2$ & $-65.3$ & $-25.3$ \\
5$^a$ & 13100 & 165 & $-237.8$ & $-79.9$ & $-39.9$ \\
\hline
\end{tabular}
\tablefoot{Measured at a range of 890\,km and scaled by $(890/350)^2$ ($+8.1$\,dB); the EIRP is independent of range. H2 is a lower bound (back-end compression). EIRP in 4\,kHz assumes isotropic radiation. The last column is the 4\,kHz EIRP relative to the SM.329 space-station limit of $-40$\,dBW in 4\,kHz for $P=20$\,dBW (Sect.~\ref{sec:sm329}); positive values exceed it. $^a$Extrapolated with the H3$\to$H4 step, not measured.}
\end{table}

In the May--June 2026 sample of \citet{indermuehle2026emlimits} (19, 8 and 5 passes at H2, H3 and H4), the range-corrected peak level is flat to about 2\,dB over the tracked elevation range, so it shows no trend with elevation over this range. Within a pass the received level varies by 10 to 20\,dB on time scales of seconds (Fig.~\ref{fig:states}), far more than the range changes, so the emission towards the telescope is not steady. They are strongly polarised. On STARLINK-11313, the raw 7860\,MHz peak is 19.0\,kJy in one polarisation and 7.5\,kJy in the other. The spectra broaden and smooth with order, from flat-topped at H2 to near-Gaussian at H4, as expected for repeated convolution. This is qualitative until equivalent bandwidths are measured. Through a 0\,dBi sidelobe at 350\,km nadir range, a single satellite exceeds the RA.769 continuum threshold \citepalias{itu2003ra769} by at least 44.4\,dB at H2, 31.7\,dB at H3 and 16.5\,dB at H4. H4 reaches the spectral-line threshold, and H2 exceeds even the VLBI threshold by at least 2.5\,dB. The extrapolated H5 lies within about 2\,dB of the continuum threshold. Section~\ref{sec:epfd} tests compliance with the radio astronomy protection criteria for the aggregate of a full constellation.

The fundamental of a tracked DTD pass (STARLINK-11163, 16\,cm, 2026-06-02) reached at least 4.3\,MJy at 850\,km \citep{indermuehle2026emlimits}, or $\geq-133.7$\,\pfdu. This is 54\,dB below the maximum pfd of $-80$\,\pfdu\ in the operator's Schedule~S \citep{spacex2023sta}. The low value results from moderate back-end compression and from the serving beam not pointing to the telescope. Tracking gives the peak level as seen by the telescope, since beam crossings are rare, but not the satellite's peak pfd. The fundamental was measured on a different satellite and receiver, so we cannot report on a harmonic-to-carrier ratio.

\subsection{Where the harmonics are generated}\label{sec:rxh2}

We computed, for each 2.07\,s integration of 211 of the 214 4\,cm observations (55 tracked DTD passes, 112 tracked passes of other satellites, 42 calibrator scans and two other scans; three passes of GPS satellites have no pointing record; 31\,158 integrations), the angular distance between the pointing, interpolated from the track file, and every DTD satellite (STARLINK-11xxx) in the nearest-epoch orbital elements. Element ages were 0.5--0.9\,d, and along-track errors of a few seconds move a satellite by several degrees. We therefore used the closest approach of any DTD satellite within $\pm10$\,s of each integration, a lower bound on the true separation; a $\pm30$\,s window gives the same result. For each integration and harmonic we measured the zone excess, the mean $r$ over the zone core (Sect.~\ref{sec:obs}) minus the mean over two side bands 30--60\,MHz either side of the zone (5180--5220 and 5280--5320\,MHz for H2), with $\sigma$ its radiometer noise from the channel scatter in the side bands. We consider a harmonic detected if the excess exceeds $5\sigma$, and the spectrum across the zone has a Pearson correlation $r>0.6$ with a  zone profile measured on the tracked DTD passes (Fig.~\ref{fig:harmim}). The shape test rejects other emitters and baseline ripples, which can raise the zone core without the spectral shape of a harmonic.

\begin{figure}
\centering
\includegraphics[width=\columnwidth]{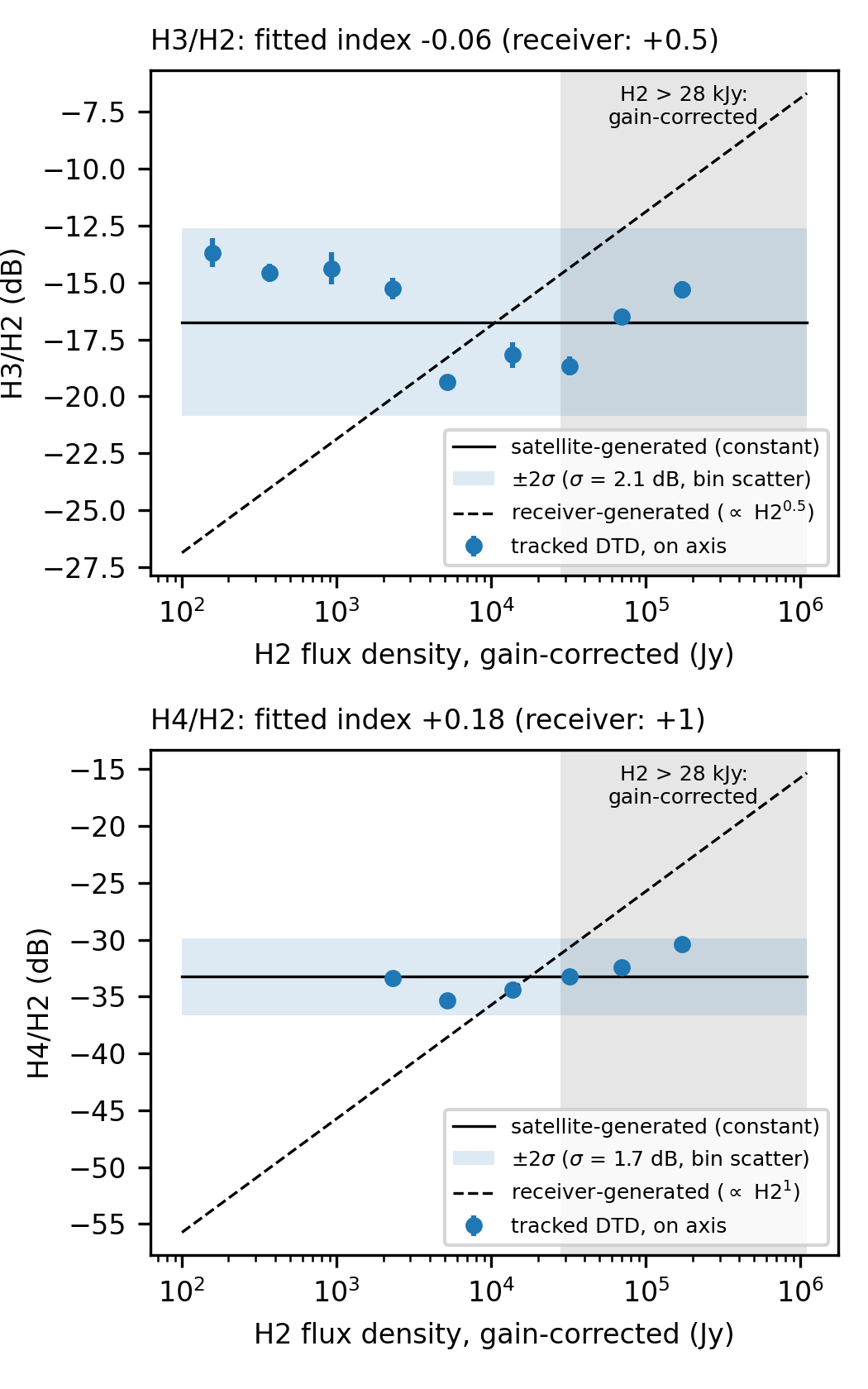}
\caption{H3/H2 (top) and H4/H2 (bottom) against the H2 flux density on tracked DTD passes, on axis (closest DTD satellite within 0.1$^\circ$), in bins of H2; ratios of the mean flux densities per bin. Above 28\,kJy (grey) H2 is corrected for back-end compression with the noise diode; integrations with a diode gain below 0.5 are excluded. Solid line and band: constant ratio expected for satellite-generated harmonics, with $\pm2\sigma$ of the bin scatter. Dashed: ratio expected if the harmonics were generated in the receiver from the received carrier. H4/H2 is shown for H2 above 0.8\,kJy, where H4 is detected.}
\label{fig:ratios}
\end{figure}

On tracked DTD passes the carrier enters through the main beam, where harmonics generated in the receiver would be strongest. The ratios of the harmonics test this. The harmonics on tracked DTD passes are generated on the satellite. If they were generated in the receiver from the received carrier power $P$, then H2\,$\propto P^2$, H3\,$\propto P^3$ and H4\,$\propto P^4$, so H3/H2 would rise as H2$^{0.5}$ and H4/H2 as H2$^{1}$. Satellite-generated harmonics instead all scale in the same way with range and with the telescope gain towards the satellite, so on axis, where the gains at the harmonic frequencies change together, their ratios are constant. Above 28\,kJy we corrected H2 for back-end compression using the noise diode, which is switched in every integration; only the H2 band compresses, by 0.3\,dB at 30--100\,kJy and 1.7\,dB above 100\,kJy. On axis, the bin means of H3/H2 show no trend with level: $-16.7$\,dB on average (bin scatter 2.1\,dB) from 100\,Jy to 280\,kJy of H2, with a fitted power-law index of $-0.06$ (95\% interval $-0.27$ to $+0.09$ from a bootstrap over passes) against $+0.5$ for receiver generation (Fig.~\ref{fig:ratios}). H4/H2 averages $-33.3$\,dB (scatter 1.7\,dB) from 1.4 to 280\,kJy, with an index of $+0.18$ ($-0.13$ to $+0.33$) against $+1$. Within single passes the decoupling is stronger still: H3 varies as H2$^{0.27\pm0.12}$ and H4 as H2$^{0.70\pm0.44}$ (21 and 9 passes with at least ten integrations), against H2$^{1.5}$ and H2$^{2}$ for products of one fundamental in the receiver. Integrations in which the diode gain falls below 0.5 (15 of 3457) are excluded, since the recorded H2 is then a ceiling. At the highest levels both ratios rise together by about 3\,dB while H4/H3 stays constant, which indicates residual compression of H2, not receiver products. A receiver-generated share of H2 would grow in proportion to H2 and hence lower H3/H2 steadily with level; instead, H3/H2 is about 1\,dB higher at 50--280\,kJy than at 3--50\,kJy. The steps between the lowest bins reflect the different passes that populate them.

\begin{figure}
\centering
\includegraphics[width=\columnwidth]{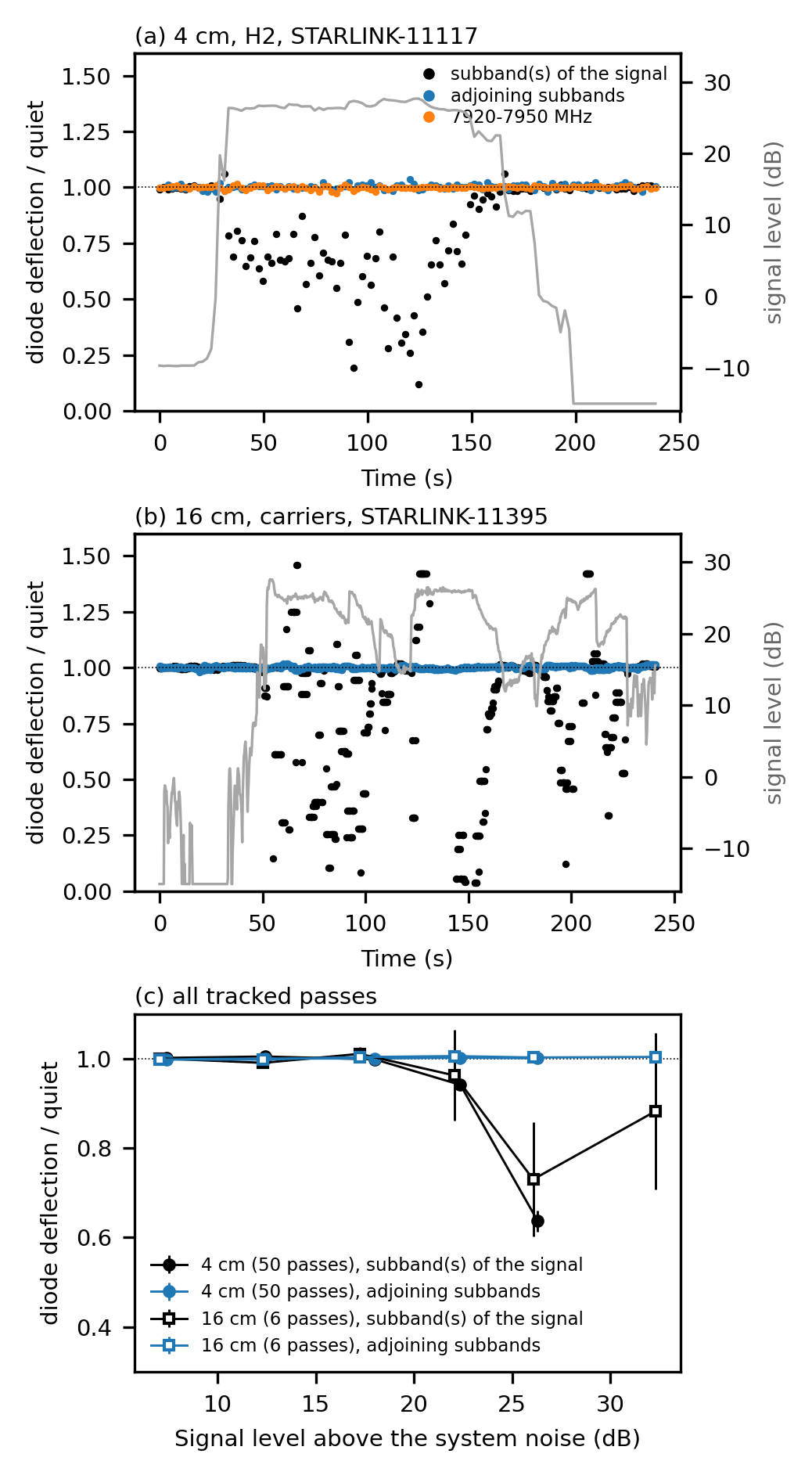}
\caption{Deflection of the switched noise diode during tracked DTD passes, relative to the part of the pass before the satellite enters the beam, measured in bands free of emission. (a) 4\,cm, STARLINK-11117 (CA02): the two back-end subbands that hold the H2 zone (black), the adjoining subbands (blue) and 7920--7950\,MHz (orange); grey: H2 level above the system noise (right axis). (b) 16\,cm, STARLINK-11395 (CA05): the subband that holds the carriers (black) and the adjoining subbands (blue); grey: carrier level. (c) Median deflection against the signal level for all tracked passes, 50 at 4\,cm (circles) and 6 at 16\,cm (squares), in the subbands that hold the signal (black) and the adjoining ones (blue).}
\label{fig:diode}
\end{figure}

The noise diode gives an independent test (Fig.~\ref{fig:diode}). If the fundamental overdrove the 4\,cm receiver, its gain would drop at all frequencies while a satellite is tracked. We measured the deflection of the switched noise diode in bands free of emission, relative to the part of the same pass before the satellite enters the beam (50 passes). In the two back-end subbands that hold the H2 zone the deflection drops once H2 exceeds about 20\,dB above the system noise, by 0.2\,dB at 20--25\,dB and 1.5\,dB at 25--30\,dB. In the adjoining subbands, at 6.4\,GHz, and next to the H3 and H4 zones, it stays within 0.6\% of its quiet value in the same integrations. The same holds at 16\,cm, where the carriers lie inside the receiver band: on six tracked passes with CA05 in December 2025 the subband that holds the carriers compresses, by 1.3\,dB at 25--30\,dB, while the seven other subbands recorded stay within 2\% with the carriers more than 30\,dB above the system noise. The compression is thus confined to the back-end subbands that contain a strong signal, as expected from the requantisation of each subband after the filterbank (J.~Tuthill, priv. comm.), and neither receiver is overdriven by the DTD signal.

The analysis based on the noise diode's signal is also compatible with the technical properties of the receiver's low-noise amplifiers. From the measurements with the 16\,cm receiver, a lower bound for the the DTD carrier is determined, as the signal is received through the sidelobes with unknown gain and the backend was saturated. The measured 4.3\,MJy on a tracked pass (Sect.~\ref{sec:res:harm}) refers to about $-74$\,dBm per polarisation at the focus of a 22\,m antenna with an aperture efficiency of at most 50\% over the 9\,MHz occupied by the carriers. The 4\,cm feed rejects frequencies at 2.6\,GHz by an estimated 20 to 30\,dB (A.~Dunning, priv. comm.). The low-noise amplifier has an input compression point of about $-30$\,dBm and second- and third-order input intercept points of about $-15$ and $-20$\,dBm. The subsequent filters suppress 2.6\,GHz by more than 100\,dB before the samplers. At the lower limit the second harmonic generated in the amplifier is below $-175$\,dBm referred to its input, 80\,dB under the $-92$\,dBm that the measured H2 of 28\,kJy delivers. At the served-cell pfd of $-80$\,\pfdu\ filed by the operator, 54\,dB above the lower limit, the amplifier would instead generate H2 at $-65$ to $-85$\,dBm, bracketing the measured range of $-80$ to $-92$\,dBm. The noise diode provides information on the actual limit: a gain change below 0.01\,dB in the adjoining subbands (Fig.~\ref{fig:diode}) limits the 2.6\,GHz power at the amplifier to below about $-50$\,dBm, so the telescope was not inside a serving beam, and an amplifier harmonic would stay below about $-85$\,dBm. The budget therefore leaves an amplifier origin open for the weaker passes; what excludes it is the behaviour of the ratios above, which no product of one fundamental reproduces.

\begin{figure}
\centering
\includegraphics[width=\columnwidth]{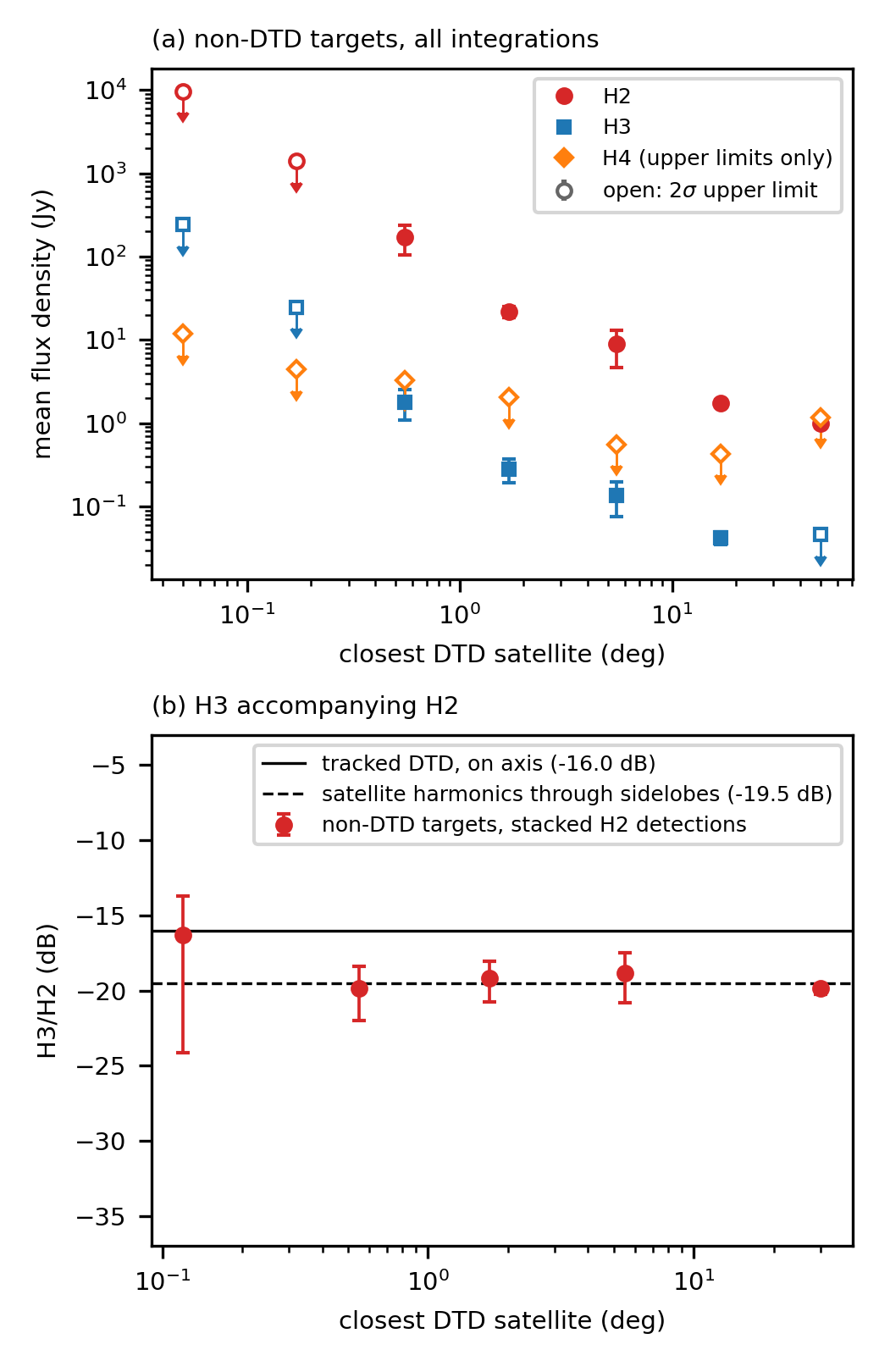}
\caption{Harmonics in scans with non-DTD targets at 4\,cm. (a) Mean flux density of H2 (circles), H3 (squares) and H4 (diamonds) against the angular distance to the closest DTD satellite within $\pm10$\,s, all integrations; open symbols with arrows: $2\sigma$ upper limits, the mean plus twice its standard error, where the mean is below twice its error. H3 and H4 are corrected for the zero level of the integrations without H2 detection. (b) H3/H2 of the stacked H2 detections against the same distance, with $1\sigma$ errors, corrected for the same zero level; solid line: tracked DTD passes on axis; dashed line: the same harmonics received through sidelobes of equal gain at both frequencies.}
\label{fig:rxh2}
\end{figure}

In scans with non-DTD targets, DTD satellites are received through the sidelobes of the dish, and several can contribute together. Figure~\ref{fig:rxh2}a shows the mean flux density of all integrations against the separation from the closest DTD satellite. The mean H2 falls from 171\,Jy at 0.3--1$^\circ$ to 22\,Jy at 1--3$^\circ$ and 1.8\,Jy at 10--30$^\circ$, and stays at 1.0\,Jy beyond 30$^\circ$. H3 follows about 20\,dB lower, from 1.8\,Jy to 0.04\,Jy; H4 is not detected.\footnote{For comparison, the radiometer noise per spectral channel (1$\sigma$) in these three frequency ranges is about 0.07\,Jy at H2, 0.03\,Jy at H3 and 0.02\,Jy at H4, noting that the features occupy many spectral channels at once.} H2 is thus present far from any DTD satellite: it is detected in 32\% of the 16\,742 integrations with no DTD satellite within 10$^\circ$, at a median of 4.3\,Jy and at most 243\,Jy. The detected fraction depends on the criterion and on the baseline estimate; the mean over all these integrations (below) does not. H3 is detected individually in 1.0\% of these integrations, at a median of 0.85\,Jy, and H2 accompanies it in 99\% of them.

Figure~\ref{fig:rxh2}b displays the H3/H2 ratio of this sidelobe H2. The H3 zone excess has a zero level of $-0.10$ to $-0.13$\,Jy in the integrations without H2 detection, which we subtracted; the stacked far H3 is significant only after this subtraction. Stacked over the H2 detections, H3/H2 is $-19.8$, $-19.2$, $-18.9$ and $-19.9\pm0.4$\,dB at 0.3--1$^\circ$, 1--3$^\circ$, 3--10$^\circ$ and beyond 10$^\circ$, 2.9 to 3.9\,dB below the on-axis value of $-16.0$\,dB. This offset is expected for the same harmonics received through sidelobes: the main-beam gain grows with frequency while the sidelobe gain does not, which lowers the ratio by 3.5\,dB. The ratio stays constant while the mean H2 of the stacks falls by 14\,dB. The far H2 follows the DTD satellites in view: in a scan with no DTD satellite within 29$^\circ$ (0823-500 OFF, 2026-06-26), H2 reaches 40\,Jy and tracks the sum over the satellites in view weighted by a 60$^\circ$ response about the pointing (Spearman $\rho=0.83$), and over 108 such scans this correlation is positive in 84\%. Emission from 5\,GHz wireless networks in 5170--5250\,MHz, in 20\,MHz channels at up to 3\,kJy, is present in 28\% of these integrations and overlaps the lower side band used as the H2 baseline. Excluding the affected integrations leaves the result unchanged: H2 is then detected in 36\% of the integrations instead of 32\%, and H3/H2 is $-18.3$\,dB instead of $-19.9$\,dB.

The H2 seen far from any DTD satellite is an aggregated effect of all DTD satellites above the horizon, received through the telescope sidelobes. Its average H3/H2 ratio is that expected for satellite harmonics received through sidelobes. The terrestrial LTE at 2630--2670\,MHz, stronger in the sidelobes than the DTD carriers, shows no comparable second harmonic at 4\,cm. 

\section{Comparison with unwanted-emission limits}\label{sec:sm329}

\textit{Emission next to the carriers} falls in the out-of-band (OoB) domain. For a multicarrier satellite, the necessary bandwidth $B_\mathrm{N}$ is the lesser of the 3\,dB transponder bandwidth and the total assigned bandwidth, and the OoB domain starts at the edges of the total assigned band \citepalias[Sect.~2.3.2]{itu2024sm1541}. Assuming the transmitter bandwidth is at least 10\,MHz, $B_\mathrm{N}=10$\,MHz and the OoB domain is 2600--2620 and 2630--2650\,MHz, the normal case for 1--3\,GHz \citepalias[Table~2]{itu2024sm1539}. The generic mask for MSS space stations, which is also that for FSS space stations, is an attenuation of $40\log_{10}(F/50+1)$\,dB in 4\,kHz below the maximum in-band power spectral density (dBsd), with $F$ the offset from the edge of the assigned band in per cent of $B_\mathrm{N}$ \citepalias[Annex~5, Section~3]{itu2024sm1541}. SM.1541 presents these masks as generic ``safety net'' limits, the least restrictive in use as national regulation, and describes the space-station mask as an upper bound that may not be applicable in a detailed examination of adjacent-band compatibility \citepalias[\textit{recommends}~4; Annex~5, Section~3]{itu2024sm1541}; \citetalias[Section~7.1.2]{ecc2025rep363} finds it conservative compared with the out-of-band emission of actual systems. Real systems are expected to attenuate more than the mask requires, and our limit lies 16\,dB below it. The maximum in-block density is about 2.3\,dB above the mean. The mask thus requires 7\,dB at 2617.5\,MHz and 19\,dB at 2610\,MHz. Our upper limit of $-33.1$\,dBc relative to the mean in-block density (Sect.~\ref{sec:res:nearim}), $-35.4$\,dBsd, lies 16\,dB below the mask at 2610\,MHz and further below it elsewhere in 2610--2620\,MHz.

Adjacent-channel emission levels for the DTD downlink were submitted to ITU-R Working Party~4C under agenda item~1.13 and are reproduced in its chairman's report \citep{itu2026wp4c35th}. Scaled to 2.6\,GHz, they correspond to an attenuation of 35, 53 and 63\,dB in the first, second and third adjacent 5\,MHz channel. Our upper limit of 33\,dB in the first two channels is consistent with these values and does not test them. The SM.1541 mask requires only 7, 16 and 22\,dB at the centres of these channels for $B_\mathrm{N}=10$\,MHz.

For scale, emission at our upper limit in a served cell at $-80$\,\pfdu\ would reach at most about $-113$\,\pfdu. This lies between the aggregate out-of-band limits for all D2D space stations of $-108$\,\pfdu\ in the UK 2.6\,GHz band \citep{ofcom2025d2dschedule} and $-120$\,\pfdu\ in the US rules for the bands below 2\,GHz \citepalias{fcc2024scs}.

\textit{Harmonics} fall in the spurious domain. SM.329 Table~2 limits space-station spurious emissions to an attenuation of $43+10\log_{10}P$ or 60\,dBc, whichever is less stringent, in 4\,kHz, with $P$ the power supplied to the antenna transmission line; for space services these are design limits \citepalias{itu2024sm329}. The Schedule~S beams combine EIRP of 49--58\,dBW with peak gains of 29--38\,dBi \citep{spacex2023sta}, so $P=20$\,dBW per beam (a 2023 filing for 1990--1995\,MHz, taken to hold at 2.6\,GHz) and the limit is $-40$\,dBW in 4\,kHz. We take the isotropic-equivalent EIRP in 4\,kHz as the radiated harmonic power, which holds if the harmonics are isotropic in all directions; we measured this only over the tracked elevation range. Then H2 exceeds the limit by at least 1.1\,dB, or 3.9\,dB with the 2.8\,dB compression estimate, while H3 and H4 are 10.7 and 25.3\,dB below it (Table~\ref{tab:harmonics}). A payload that drives several beams from one array has a larger total $P$, which relaxes the 60\,dBc term. H2 is thus near or above the limit.

These limits refer to the antenna port. The carrier gains 29--38\,dBi from the array, whereas the harmonics are received at these levels well outside the serving beams. Studies that model harmonics as a fixed number of dB below the carrier, using the carrier's antenna pattern, therefore underestimate them outside the serving beam. SM.329 recognises that for active antennas the power at the antenna line misses emissions created within the antenna, and that field strength or pfd at a distance should then be established \citepalias[\textit{considering}~l;][]{itu2024sm329}. Measurements such as these provide that quantity.

\subsection{Back-off and other mitigations}\label{sec:mitigation}

Output back-off, operating the power amplifier a number of dB below its saturated output power, makes it more linear: each dB of back-off lowers the carrier by 1\,dB and an order-$N$ product by $N$\,dB. A total-power-limited amplifier that adds carriers lowers the power per carrier, and with it each order-$N$ product by $3N$\,dB per halving. \citetalias{ecc2011rep171} suggested a 3--5\,dB power reduction for Iridium and showed that channel planning can keep the products out of the radio astronomy band. A filter after the amplifier attenuates the harmonics before they reach the array. The array does not confine the harmonics to the serving beams: they are received at the measured levels in directions far from them (Sect.~\ref{sec:res:harm}). Digital predistortion can lower low-order products.

\section{Aggregate impact on radio astronomy}\label{sec:epfd}
Based on the flux densities determined for the harmonics and on the out-of-band mask, it is possible to assess the theoretical impact of these signals on radio astronomy observations. Depending on the chosen carrier frequencies, which vary in different regions of the world, the out-of-band features or harmonics could overlap radio astronomy bands with primary allocations. As examples, we analyse three generic cases and one planned deployment:
\begin{enumerate}
    \item \textit{Out-of-band case:} If a DTD carrier block is operated with the primary passive RAS band at 2690--2700\,MHz in its third and fourth adjacent channels, out-of-band emission at the level that the SM.1541 mask allows there (19 to 28\,dB below the in-band density, Sect.~\ref{sec:sm329}) would, in a served cell at $-80$\,\pfdu, put $-99$ to $-108$\,\pfdu\ into the radio astronomy band. We use the lower end of this range, $-108$\,\pfdu, the case most favourable to the operator, converted to an EIRP at the 350\,km nadir range. This case is hypothetical: next to the 2620--2630\,MHz block we measure an upper limit 33\,dB below the carrier density (Sect.~\ref{sec:res:nearim}), at least 5\,dB below the level assumed here;
    \item \textit{Second harmonic case:} The primary RAS band at 4990--5000\,MHz could be subject to second harmonics with an EIRP of $\geq -38.9$\,dBW in 4\,kHz from a DTD carrier at 2497.5\,MHz, in the band 2483.5--2500\,MHz allocated to the mobile-satellite service (space-to-Earth). This case is illustrative only, as no DTD operator currently plans to use this band in any region. Among the DTD bands authorised to date \citepalias{fcc2024da1193}, the second-order sum product of carriers in 2345--2360 and 2620--2690\,MHz reaches the same RAS band, at a level comparable to H2 if both carriers share a transmit chain;
    \item \textit{Fifth harmonic case:} The primary RAS band at 10600--10700\,MHz could be subject to fifth harmonics with an EIRP of $\approx -80$\,dBW in 4\,kHz from one or more DTD carriers somewhere in the range 2120--2140\,MHz.
    \item \textit{Planned deployment below 2690--2700\,MHz:} In Australia the mobile network operator Optus holds 2670--2690\,MHz in this band \citepalias{acma2013auction} and has indicated its intention to use the band for DTD services with the same constellation \citepalias[p.~28]{acma2025esl4}.. We take a downlink at 2670--2680\,MHz, the lower half of that block; the passive band then occupies its third and fourth adjacent channels. A downlink in the upper half would place the passive band in the first and second adjacent channels, where the mask allows 7 to 16\,dB and the operator has submitted 35 and 53\,dB. We evaluate this case with the DTD satellites in orbit and for three emission levels in the passive band: the SM.1541 mask as in the out-of-band case (28\,dB below the in-band density), the upper limit that we measure in the same channels below the 2620--2630\,MHz block (45\,dB, Sect.~\ref{sec:res:nearim}), and the attenuation of 63\,dB submitted by the operator for the third adjacent channel \citepalias[Annex 10]{itu2026wp4c35th}.
\end{enumerate}

In all cases the EIRP spectral density is assumed to fill the whole radio astronomy band. The EIRPs are peak values, both over the harmonic zone and over the pass: on axis, the range-corrected H2 of the tracked passes has a median of 2\,Jy and a mean of 9\,kJy over all integrations, 15\,dB below the peak used here, with 55\% of the integrations below 10\,Jy and 17\% above 10\,kJy (Sect.~\ref{sec:rxh2}). The harmonic cases thus describe the aggregate when every satellite radiates at its peak; with the time-mean level the H2 and H5 margins are 15\,dB smaller. The fifth harmonic case requires 2120--2140\,MHz to be fully occupied by carriers. Averaged over the radio astronomy band, the convolution model gives levels about 3\,dB (H2 of a single 5\,MHz carrier) and 2\,dB (H5) below the peak; for H2 this is partly offset by the back-end compression of the measured value.

The aggregate power, injected by all satellites of a constellation into a radio astronomy receiver, can be inferred using the equivalent power flux density (epfd) method, defined in \citetalias{itu2007m1583}, \citetalias{itu2007s1586} and \citetalias{itu2015ra1513}. \citet{divruno2023lofar} and \citet{indermuehle2026emlimits} apply the method to measured satellite emissions. Therefore, we omit a detailed description of the epfd method. We express the epfd as the power flux density that would deliver the aggregate received power to an antenna of 0\,dBi gain, the reference used for the RA.769 thresholds.

At the time of writing, 638 DTD Starlink satellites are in orbit \citep{mcdowell2026starlink}, and many more are planned. While concrete plans of satellite numbers and orbit configurations change rapidly, we make the assumption of two orbital shells with 5280 satellites (48 orbital planes with 110 satellites each) at 340\,km altitude and 3360 satellites (28 orbital planes with 120 satellites each) at 525\,km, both with an inclination of 53$^\circ$. For the planned deployment we instead used the orbital elements of the 642 DTD satellites in orbit on 1 June 2026, about half at an inclination of 43$^\circ$ and half at 53$^\circ$, at altitudes near 357\,km, again all transmitting continuously. For simplicity, it is assumed that all satellites transmit continuously. In reality, a fraction of them may be switched off, which lowers the aggregate by 3\,dB for a duty cycle of 50\%, and the harmonics are intermittent on time scales of seconds (Sect.~\ref{sec:res:harmim}), which is the larger effect: the time-mean level is 15\,dB below the peak, as stated above.

We ran the simulations for two telescopes: the Effelsberg 100-m radio telescope in Germany, which has receivers for all three generic cases and, as one of few 100-m class telescopes, is often used for spectrum management compatibility studies, and a 22-m antenna of the ATCA, where the observations were made. The protection criteria for the studied radio astronomy bands are listed in Table~1 of \citepalias{itu2003ra769}. The threshold values in the table refer to 10\% of the rms noise level that a typical low-noise receiver system will achieve after an integration time of 2000\,s. These levels must not be exceeded for more than 2\% of the epfd samples \citepalias{itu2015ra1513}. As usual in spectrum management compatibility calculations, we used the antenna pattern of \citetalias{itu2003ra1631} for the radio astronomy telescope. Although this is a fairly simple model, it is well suited for aggregation calculations.

For each case, 400 iterations of the epfd simulations were performed, each with a different initial phase of the orbital configuration (spread over one day). As the SNIFFLES campaign revealed, the peak harmonic levels show no trend with elevation over the tracked range. As a simple approximation it is assumed that the signals are transmitted isotropically. The operator has stated in ITU-R Working Party 4C that unwanted emission in the out-of-band domain radiates with the pattern of a single element of the array, because the non-linear products of many beams lose the spatial correlation that forms the beams \citepalias[Annex 10, Sect.~3.1.1 and Appendix~1]{itu2026wp4c35th}; this is contested in the same document, and it concerns the out-of-band domain rather than harmonics. Such an element pattern can be approximated by the cosine of the nadir separation angle. Therefore, for comparison, we also ran the calculations under this assumption. This pattern lowers the 98th-percentile epfd by only 1.4 to 2.3\,dB (Table~\ref{tab:epfd}), since the aggregate is dominated by satellites at high elevation, close to the nadir direction of the satellite. The time resolution in our simulations was set to 0.25\,s. 

\begin{figure}
\centering
\includegraphics[width=\columnwidth]{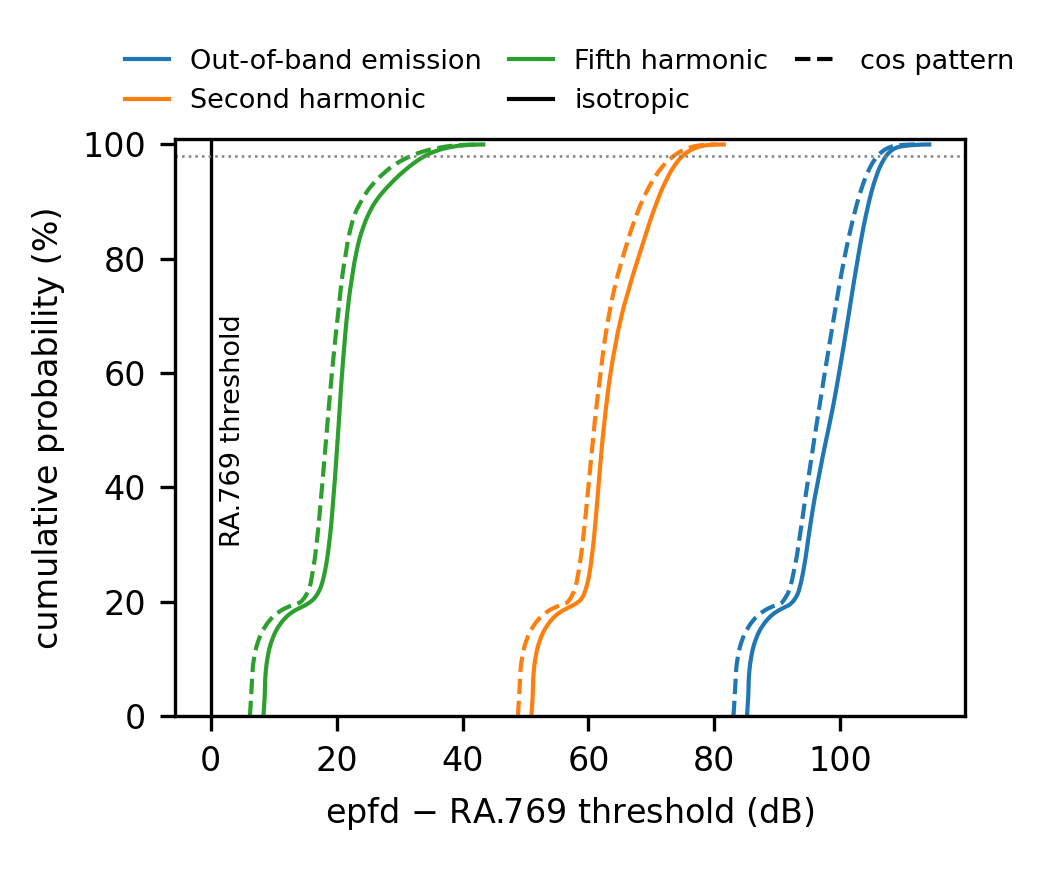}
\caption{Cumulative distribution of the epfd at the Effelsberg 100-m telescope relative to the RA.769 continuum threshold, for the three generic cases of Sect.~\ref{sec:epfd}, pooled over 400 iterations and all 3$^\circ$ sky cells. Solid: isotropic emission; dashed: single-element pattern $\cos\theta_\mathrm{nadir}$. Dotted: 98\% level.}
\label{fig:epfd}
\end{figure}

\begin{table*}
\caption{Epfd of DTD satellites at the Effelsberg 100-m telescope and at a 22-m antenna of the ATCA: 400 iterations of 2000\,s at 0.25\,s resolution, pointings down to the lowest elevation of each telescope (8$^\circ$ at Effelsberg, 12$^\circ$ at the ATCA), RA.1631 antenna pattern.}
\label{tab:epfd}
\centering
\small
\begin{tabularx}{\textwidth}{Xlrrrr}
\hline\hline
Case & RAS band & EIRP$_{4\,\mathrm{kHz}}$ & RA.769 & \multicolumn{2}{c}{Margin (dB)} \\
 & (MHz) & (dBW) & \multicolumn{1}{c}{(dB(W\,m$^{-2}$))} & isotropic & cos \\
\hline
\raggedright\textit{Effelsberg, planned constellation of 8640 satellites} & & & & & \tabularnewline
Out-of-band emission & 2690--2700 & $-10.1$ & $-176.6$ & $-107.3$ & $-105.9$ \\
Second harmonic & 4990--5000 & $-39.0$ & $-171.3$ & $-75.0$ & $-73.6$ \\
Fifth harmonic & 10\,600--10\,700 & $-80.0$ & $-159.7$ & $-34.1$ & $-31.8$ \\
\hline
\raggedright\textit{ATCA, planned constellation of 8640 satellites} & & & & & \tabularnewline
Out-of-band emission & 2690--2700 & $-10.1$ & $-176.6$ & $-99.9$ & $-98.1$ \\
Second harmonic & 4990--5000 & $-39.0$ & $-171.3$ & $-66.8$ & $-65.3$ \\
Fifth harmonic & 10\,600--10\,700 & $-80.0$ & $-159.7$ & $-26.6$ & $-25.1$ \\
\hline
\raggedright\textit{ATCA, the 642 DTD satellites in orbit on 1 June 2026; downlink at 2670--2680\,MHz; emission in the third and fourth adjacent channels, 2690--2700\,MHz, at:} & & & & & \tabularnewline
Mask level (28\,dB) & 2690--2700 & $-10.1$ & $-176.6$ & $-95.3$ & $-93.6$ \\
Measured upper limit (47\,dB) & 2690--2700 & $-29.3$ & $-176.6$ & $-76.1$ & $-74.5$ \\
Submitted by the operator (63\,dB) & 2690--2700 & $-45.1$ & $-176.6$ & $-60.3$ & $-58.6$ \\
\hline
\end{tabularx}
\tablefoot{EIRP spectral density per satellite, assumed to fill the radio astronomy band (Sect.~\ref{sec:epfd}); for the out-of-band case it is the $-108$\,\pfdu\ mask level converted at 350\,km. The harmonic EIRPs are those of Table~\ref{tab:harmonics}, rounded to $-39.0$ and $-80.0$\,dBW in the simulations. Margin: RA.769 continuum threshold minus the 98th-percentile epfd, for isotropic emission and for the $\cos\theta_\mathrm{nadir}$ element pattern; negative values exceed the threshold. The epfd is the aggregate received power expressed as the power flux density in the radio astronomy band that would deliver it to an antenna of 0\,dBi gain, the reference of the RA.769 thresholds. In the last block the levels are given in dB below the in-band power density of the carriers; the epfd scales linearly with this level, and the measured upper limit is that found in the third and fourth adjacent channels below the 2620--2630\,MHz block (2600--2610\,MHz).}
\end{table*}

All three generic cases exceed the RA.769 thresholds at both telescopes in every sky cell, a data loss of 100\% (Fig.~\ref{fig:epfd}, Table~\ref{tab:epfd}). At Effelsberg the 98th-percentile epfd exceeds the threshold by about 107\,dB for the out-of-band case (106\,dB for the cos pattern) and by 75 and 34\,dB for the second and fifth harmonics (74 and 32\,dB). At the ATCA the excess is 100, 67 and 27\,dB (98, 65 and 25\,dB), 7 to 8\,dB less; the ATCA lies at a lower latitude, further from the 53$^\circ$ inclination of the shells where the density of satellites on the sky peaks. The out-of-band case applies the served-cell pfd to every satellite in view: at the 350\,km nadir range, $-108$\,\pfdu\ in all directions is an EIRP of $+24$\,dBW in 10\,MHz per satellite, more than the 20\,dBW carrier power at the antenna port (Sect.~\ref{sec:sm329}). This is an upper bound, which holds if the out-of-band emission is not beamformed at all, as the operator claims for the out-of-band domain (beamforming efficiency 0\%). A single satellite at this pfd level already exceeds the threshold by 79\,dB, which needs none of the constellation assumptions. For the planned deployment at 2670--2680\,MHz, with only the 642 satellites now in orbit, the epfd at the ATCA exceeds the threshold of the passive band in every sky cell, at the 98th percentile by 95\,dB at the mask level, by 76\,dB at our measured upper limit, and by 60\,dB at the level submitted by the operator (94, 75 and 59\,dB for the cos pattern). The same assumption applies here: every satellite in view is taken to radiate, in all directions, the level that belongs to a served cell. The measured wings of the harmonic zones (Sect.~\ref{sec:res:harmim}) extend this to radio astronomy bands just outside a harmonic zone. Averaged over the first 9\,MHz beyond the zone edge, the products lie on average 19\,dB below the mean density in the zone centre for H2, H3 and H4, and between 15 and 23\,dB at a given moment. The simulated EIRPs are peak densities, about 2\,dB above that mean, so the wings lie about 21\,dB below the simulated levels. Since the epfd scales linearly with the EIRP, a 10\,MHz radio astronomy band within 10\,MHz of an H2 zone would still exceed its threshold by about 54\,dB. If the H5 wings, which we did not measure, follow the same ratio, a band covered by them would exceed its threshold by about 13\,dB. In the 100\,MHz band at 10.6--10.7\,GHz, a 10\,MHz wing covers a tenth of the band, which brings the band average down to about the threshold.

The epfd margins are far larger than the per-satellite margins of Sect.~\ref{sec:sm329}: a satellite that meets the ITU-R limits can still contribute to an aggregate that exceeds the RA.769 thresholds by tens of dB.

\section{Conclusions}\label{sec:conclusions}

\begin{enumerate}
\item Harmonics of the Starlink DTD downlink in Australia and intermodulation products around them were clearly detected with the ATCA, which raises concerns regarding possible interference into protected radio astronomy bands.

\item Next to the carriers no emission from the satellite is detected. The power density in 2610--2620\,MHz is below $-33$\,dBc. Features in this band that resemble intermodulation products arise in the back-end.

\item The harmonics H2 to H4 fall by about 16\,dB per order and show no trend with elevation over the tracked range. H2 ($\geq-38.9$\,dBW in 4\,kHz, isotropic-equivalent) is at or above the SM.329 space-station limit of $-40$\,dBW for a 100\,W port power. Their ratios show no trend with the received level, and within passes H3 does not follow H2 as a receiver product would, so they are generated on the satellite; H2 of a few Jy is also detected in 32\% of the integrations more than 10$^\circ$ from any DTD satellite; H3, detected when these integrations are stacked, has the ratio to H2 of satellite harmonics received through sidelobes. The noise diode shows no gain compression of the receiver during tracked passes, which limits the carrier power at the 4\,cm amplifier to below about $-50$\,dBm; the receiver origin is excluded by the ratios, not by the power budget.

\item Measured against the ITU-R limits, the upper limit next to the carriers lies 16\,dB below the out-of-band mask and H3 and H4 meet the SM.329 spurious-domain limit, while H2 is at or above it. Meeting these limits does not protect radio astronomy. The limits apply per transmitter at the antenna port, whereas the RA.769 thresholds apply to the aggregate of all satellites in view (\citetalias{ecc2025rep363}; \citealt{indermuehle2026emlimits}). A single satellite that meets SM.329 at H3 and H4 exceeds the RA.769 continuum threshold through a 0\,dBi sidelobe by 31.7 and 16.5\,dB. For a constellation of 8640 satellites, the epfd exceeds the thresholds at Effelsberg and at the ATCA, by 25 to 107\,dB at the 98th percentile (Sect.~\ref{sec:epfd}).

\item Out-of-band emission that meets the mask can still exceed the thresholds of a radio astronomy band in its adjacent channels. For the same two-carrier downlink at 2670--2680\,MHz, the band 2690--2700\,MHz, where RR No.~5.340 prohibits all emissions, is within the third and fourth adjacent channels. The mask requires only 19 to 28\,dB of attenuation there, so a downlink that just meets it in a served cell at $-80$\,\pfdu\ could put $-99$ to $-108$\,\pfdu\ into the band, 79 to 88\,dB above the RA.769 continuum threshold of $-187$\,\pfdu\ \citepalias{itu2003ra769}. For a constellation of 8640 satellites at this level, the epfd at Effelsberg exceeds the threshold by 107\,dB at the 98th percentile (Sect.~\ref{sec:epfd}), so protecting the band requires out-of-band levels about 100\,dB below the mask. Optus holds 2670--2690\,MHz in Australia and intends to use the band for DTD services with the same constellation \citepalias{acma2025esl4}; considering only the DTD satellites now in orbit, and at the attenuation of 63\,dB submitted by the operator for that offset, the epfd at the ATCA would still exceed the threshold by about 60\,dB. Terrestrial base stations in the same band are subject to their own unwanted-emission limits \citepalias{3gpp36104}, but their level at a telescope depends on distance and terrain and can be controlled by coordination zones around the site. A satellite is above the horizon of every telescope in its coverage, so its own out-of-band level must meet the threshold.

\item Inside the harmonic zones the spectra follow the $n$-fold convolution of the carrier spectrum. Just outside them, intermodulation products of order $n+2$ and higher lie on average 19\,dB below the zone at H2, H3 and H4, the same on two antennas at epochs seven months apart, although the earlier epoch rests on one dominant pass per harmonic.. At H2 their ratio to the zone switches between about $-23$ and $-15$\,dB within seconds while the profile inside the zone stays the same, and it follows neither the zone level nor H3/H2.

\item Further observations should track one satellite simultaneously at 16\,cm and 4\,cm, cover 13\,GHz for H5, and measure a satellite with carriers in more than one band.
\end{enumerate}

\begin{acknowledgements}
      We thank A.~Dunning for the characteristics of the 4\,cm receiver and J.~Tuthill for discussions of the BIGCAT filterbank and requantisers. The Australia Telescope Compact Array is part of the Australia Telescope National Facility, which is funded by the Australian Government for operation as a National Facility managed by CSIRO. We acknowledge the Gomeroi people as the Traditional Owners of the Observatory site.
\end{acknowledgements}

\bibliographystyle{aa}
\bibliography{intermods}

\end{document}